\documentclass[onecolumn]{emulateapj}
\usepackage{apjfonts}
\usepackage{graphicx}
\newcommand{\be}{\begin{equation}}
\newcommand{\ee}{\end{equation} }
\newcommand{\ba}{\begin{eqnarray}}
\newcommand{\ea}{\end{eqnarray}}
\newcommand{\nuem}{\nu_{\rm em}}
\newcommand{\tildenuem}{\tilde{\nu}_{\rm em}}
\newcommand{\bnabla}{\mbox{\boldmath$\nabla$}}
\newcommand{\bbeta}{\mbox{\boldmath$\beta$}}
\newcommand{\nn}{\mbox{} \nonumber \\ \mbox{} }

\shorttitle{Hot Electromagnetic Outflows I}
\shortauthors{Russo \& Thompson}	
\slugcomment{To be published in the Astrophysical Journal}
\begin{document}

\title{Hot Electromagnetic Outflows I:  Acceleration and Spectra}
\author{Matthew Russo}
\affil{Department of Physics, University of Toronto, 60 St. George St., Toronto, ON M5S 1A7, Canada.}
\author{Christopher Thompson}
\affil{Canadian Institute for Theoretical Astrophysics, 60 St. George St., Toronto, ON M5S 3H8, Canada.}

\begin{abstract}
The theory of cold, relativistic, magnetohydrodynamic outflows is generalized by the inclusion
of an intense radiation source.  In some contexts, such the breakout of a gamma-ray burst jet from a star, 
the outflow is heated to a high temperature at a large optical depth.  Eventually it becomes transparent
and is pushed to a higher Lorentz factor by a combination of the Lorentz force and
radiation pressure.  We obtain its profile, both inside and 
outside the fast magnetosonic critical point, when the poloidal magnetic field 
is radial and monopolar.   Most of the energy flux is carried by the radiation field and the toroidal
magnetic field that is wound up close to the rapidly rotating engine.  Although the entrained matter 
carries little energy, it couples the radiation field to the magnetic
field.  Then the fast critical point is pushed inward from infinity and,
above a critical radiation intensity, the outflow is accelerated mainly by radiation pressure.
We identify a distinct observational signature of this hybrid outflow:  a hardening
of the radiation spectrum above the peak of the seed photon distribution, 
driven by bulk Compton scattering.  The non-thermal spectrum -- obtained by a Monte Carlo method --
is most extended when the Lorentz force dominates the acceleration, and the seed photon beam is wider than the 
Lorentz cone of the MHD fluid.   This effect is a generic feature of hot, magnetized outflows interacting 
with slower relativistic material.  It may 
explain why some GRB spectra appear to peak at photon energies above the original Amati et al. scaling.
A companion paper addresses the case of jet breakout, where diverging magnetic flux 
surfaces yield strong MHD acceleration over a wider range of Lorentz factor.
\end{abstract}


\section{Introduction}
Some astrophysical sources release energy at prodigious rates, as a result of their extreme magnetism 
and/or rotation.  Gamma-ray bursts (GRBs) and soft gamma repeater flares are familiar examples, but the range 
of possibilities extends to rapidly rotating white dwarfs formed in binary mergers, or young and hot magnetars.
The `engine' generates an outflow that contains both an intense magnetic field and 
also a powerful flow of radiation.   A gamma-ray burst jet becomes hot as it works its way through a 
preceding layer of stellar material, as in the collapsar model \citep{macfadyen99}; or interacts with a powerful 
neutron-rich wind that was generated in the initial stages of a binary merger (e.g. \citealt{dessart09}).
A blackbody component is also expected to be carried outward from the engine itself even in the absence of jet interaction \citep{goodman86}.

We have previously argued that the full GRB phenomenon cannot be captured by `thermal fireballs' 
\citep{goodman86,shemi90}, or `Poynting-dominated outflows' \citep{lyutikov03} alone.  Instead both components are needed: 
thermal photons to provide seeds for gamma-ray emission, and large-scale magnetic fields to drive the outflow
and then trigger non-thermal activity and fast variability \citep{thompson94,meszaros97,drenkhahn02,thompson06,giannios07,zhang11}.
In some other circumstances, such as white dwarf merger remnants, the photon luminosity 
is bounded by the Eddington value, but the spindown luminosity can be vastly greater.  (It is, of course, 
possible to find situations in which this inequality is reversed:  the photon energy flux in giant 
magnetar flares is strong enough to pull some magnetic flux away from the star, but the total energy flux is 
probably dominated by photons.)

\subsection{Acceleration}

Our first interest here is in how such a `hot electromagnetic outflow' is accelerated.  We focus on
stationary, axisymmetric flows in the ideal MHD limit, but add a radiation field with
a prescribed source radius that can have arbitrary size with respect to the light cylinder.
The inertia of the magnetic field dominates that of the matter to which it is tied, so that
the outflow can achieve relativistic speeds.  The magnetofluid is accelerated by two mechanisms:
the Lorentz force (which operates in cold MHD winds);  and scattering off the radiation field
(which operates in thermal fireballs).  

In the optically thin regime, the radiation field is
self-collimating, so that a relativistically moving frame can be defined in which
the radiation exerts a vanishing net force on the matter.  Radiation pressure dominates matter pressure, and 
with the possible exception of a small region close to the engine, the radiation temperature
lies far below the rest energy of the advected particles.  The outflow can therefore
be assumed cold outside the transparency surface, and scattering operates in the Thomson limit.
In this first paper, we follow \cite{goldreich70} in restricting the poloidal magnetic field to a monopolar
geometry.  This limits the efficiency of MHD acceleration (the fast critical point of the cold MHD flow
sits at infinity).  A much stronger Lorentz force can arise through differential decollimation of
magnetic flux surfaces (e.g. \citealt{tchek09}),  but even then some parts of a cold MHD outflow 
may not have the requisite geometry.  A companion paper considers jet geometries and allows for 
differential decollimation of the jet with respect to the radiation field (Russo \& Thompson 2012, paper II).

The efficiency of radiative acceleration outside the fast magnetosonic point was noted by
\cite{thompson06} in the context of GRBs.
Previous work by \cite{li92} and \cite{beskin04} considered the interaction of a relativistic MHD outflow 
with a quasi-isotropic radiation field.  In a first approximation, the outflowing matter feels a net
drag force.  \cite{beskin04} focused on a slightly decollimating outflow with fast MHD acceleration, and
analysed the influence of radiation drag in integral form through the changes imparted to the energy and
angular momentum.  They considered changes in the shape of the magnetic surfaces, and therefore in the MHD
acceleration rate, imparted by the radiation field, but the first-order effect addressed in this paper -- 
the outward acceleration due to the strong radial anisotropy of the radiation field -- was absent in their calculations.  

We focus on time-independent outflows, because acceleration by radiation pressure becomes important 
before effects associated with the radial structure of the flow.
The acceleration of a magnetized slab in planar geometry was studied numerically by \cite{granot11}.  The slab is initially static; its mean Lorentz factor quickly reaches $\sigma^{1/3}_0$, where $\sigma_0$ 
is the initial magnetization, and then continues to grow.  This effect may be 
relevant to GRBs after a jet has made a transition to a planar geometry at a distance 
$r \ga c\Delta t\sim 3\times 10^{11} (\Delta t/10~{\rm s})$ cm, where $\Delta t$ 
is the duration of the prompt phase.  If the shell is already moving relativistically, only the outermost
fraction $\sim 1/2\Gamma^2$ of the shell will experience this type of acceleration.  The interaction of
this thin layer with an external plasma shell, which absorbs momentum from the magnetic field 
(e.g. \citealt{thompson06}), must also be taken into account.  Typically the bulk of the shell
receives momentum from the radiation field before it comes into causal contact.  

We do not address how the outflow might accelerate below its scattering photosphere.  
For example, periodic reversals of a toroidal field might induce magnetic reconnection, 
break the degeneracy between magnetic pressure gradient and curvature forces, 
and thereby induce a net outward acceleration \citep{drenkhahn02}.   The
high efficiency claimed for this mechanism depends on neglecting the input of enthalpy,
and therefore inertia, into the magnetofluid as the magnetic field reconnects.  When an 
enthalpy source term is included, one only obtains relativistic bulk motion if a high 
fraction of the unsigned magnetic flux is removed by reconnection.  This requires, at 
a minimum, that the net (signed) magnetic flux carried by the outflow is small compared
with the unsigned flux, and that the non-radial magnetic field maintains a strictly uniform
(e.g. toroidal) direction across current sheets.  Our focus here and in paper II is 
therefore on ideal MHD effects.

\subsection{Non-thermal Spectrum}

The second focus of this paper is on the non-thermal spectrum of photons that are scattered by the
outflowing matter.  We ignore any effects of internal dissipation in the outflow, and focus purely on
the spectral signature of bulk relativistic motion.  We find a broad and flat extension of the seed
spectrum when the outflow is rapidly accelerated by the Lorentz force, so that 
the seed photon beam is wider than the Lorentz cone of the MHD outflow.  The smoothness of the
scattered spectrum -- in particular, the presence or absence of a residual bump at the seed
thermal peak -- is shown to depend on the net optical depth that is seen by the broader, unscattered
photon beam.

Existing calculations of non-thermal `photospheric' emission from GRB outflows
\citep{giannios06,beloborodov11} generally assume that the Lorentz
factor of the outflow has saturated at the transparency surface, as do calculations of 
the low-frequency spectral tail of thermal photospheres observed at oblique angles (e.g. \citealt{peer08,lazzati11}).  
We broaden this approach by noting that locking between the advected thermal photons and the relativistic outflow will be broken 
in the presence of a somewhat slower component of the outflow.   For example, material with a Lorentz factor $\sim 3-5$ times smaller
than the magnetofluid would be present as the result of the interaction of a jet with a star \citep{thompson06}.  In a neutron star
binary merger, a relativistic magnetofluid could interact with a sub-relativistic, neutron-rich wind \citep{bucciantini12}.  The 
slower material scatters advected X-ray photons into a broader beam, which continues to interact strongly with the faster magnetofluid
even at low optical depth.  It has long been realized that relativistic material moving into an isotropic bath of very low-frequency
(optical-UV) photons will upscatter them as it {\it loses} energy to Compton drag (e.g. \citealt{sikora94,ghisellini00}).   In that case a
spectral slope $F_\nu \sim \nu^{-1/2}$ is generated by the decaying peak energy of the upscattered photons -- a different effect from that
considered here, and a spectrum somewhat softer than that observed in the low-energy tails of GRBs.

\subsection{Plan}

The plan of the paper is as follows.  Section \ref{s:postfast} explains
the efficiency of radiation-driven acceleration beyond the fast magnetosonic 
point, and explains further our critique of acceleration by magnetic reconnection.  
The relativistic wind equations including radiation pressure
are described in Section \ref{s:eom}, and in Section \ref{s:mono} they
are specialized to a monopole magnetic field.  The flow
properties near the fast critical point are analysed and the numerical
method described.  Numerical results are presented in Section \ref{s:results}.
The spectrum of scattered photons 
is calculated in Section \ref{s:spectrum} by a Monte Carlo method, and the low- and high-frequency
components of the spectrum discussed in the context of GRBs.
Our results are summarized in Section \ref{s:conclusions}, and
the effect of rotation in the photon source is briefly discussed in 
Appendix \ref{s:rotation}.

\section{Acceleration of Magnetically Dominated Flows by Radiation Pressure}\label{s:postfast}

The outflow may be divided into an inner, optically thick part, in which the radiation field is
effectively tied to the matter; and an outer part through which the radiation can flow 
almost unimpeded.\footnote{A hard radiation spectrum, extending above $m_ec^2$ in the frame
of the ambient material, can trigger an $e^+e^-$ cascade outside the photosphere of a 
{\it relativistic} outflow \citep{tm00,beloborodov02}.  We ignore the effect of pair creation on the optical depth in
this paper.}  Within this outer transparent zone, the radiation field becomes progressively
more collimated with increasing distance from the engine.  Even at the transparency radius,
it can remain so intense as to push matter to much higher Lorentz factors than were achieved
by hydromagnetic stresses operating in the inner zone.  


The radiation has a positive accelerating effect outside the scattering photosphere
if the terminal Lorentz factor $\Gamma_\infty$ of the matter exceeds
the Lorentz factor $\Gamma(r_\tau)$ at the transparency radius $r_\tau$.  The intensity of the radiation
field is measured by the compactness, which here is normalized to the mean mass $\bar m$ per
scattering charge:
\be\label{eq:chitau}
\chi(r_\tau) = {\sigma_T \over \bar m c^3 r_\tau}{dL_\gamma\over d\Omega}.
\ee
Here $L_\gamma$ is the isotropic radiation luminosity, and $\sigma_T$ is the Thomson cross-section
(we only consider classical electron scattering in this paper).  Defining the scattering optical depth
\be\label{eq:tauT}
\tau_{\rm es} = \int \Gamma {\rho(r)\over\bar m}(1-\beta_r)\sigma_T dr
\ee
in a slowly rotating outflow with speed $\beta_r c$ and proper mass density $\rho$,
the surface $\tau_{\rm es} = 1$ can be related to the flux of rest mass $\dot M$, 
\be\label{eq:rtau}
r_\tau \simeq {\sigma_T \over 6\Gamma^2(r_\tau)\,\bar m c} {d\dot M\over d\Omega}.
\ee
The magnetization of a relativistic outflow can be expressed in terms of the ratio of Poynting flux
to rest energy flux, which is, far outside the speed of light cylinder,
\be
{d L_{\rm P}\over d\Omega} \sim {B_\phi^2\over 4\pi} r^2 c \simeq \sigma {d\dot M\over d\Omega} c^2.
\ee
The magnitude of $\sigma$ in GRB outflows is unknown, but there are a number of reasons to expect
a range of values within different components of the same outflow.   The simplest case -- but perhaps not
the most relevant for classical GRBs -- involves a rapidly rotating nascent magnetar, whose wind 
can achieve a magnetization $\sim10^3$ at late times after an early dirty wind phase \citep{metzger11}.

At $r = r_\tau$, one therefore has the relation between photon compactness and magnetization,
\be
\chi(r_\tau) \sim 6\Gamma^2(r_\tau) {dL_\gamma/d\Omega\over dL_{\rm P}/d\Omega} \sigma.
\ee
The terminal Lorentz factor of a radiation-driven baryonic wind is (see Section \ref{s:dispphot})
\be
\Gamma_\infty(r_\tau) \sim [\chi(r_\tau) \Gamma(r_\tau)]^{1/4} = 
      \left[6\Gamma^3(r_\tau){dL_\gamma/d\Omega\over dL_{\rm P}/d\Omega} \sigma\right]^{1/4}.
\ee
Hence the outflow emerges into transparency below this limiting Lorentz factor if
\be
{\Gamma(r_\tau)\over \Gamma_\infty} = \left[{\Gamma(r_\tau)\over 6(L_\gamma/L_{\rm P})\sigma}\right]^{1/4} < 1,
\ee
corresponding to $\Gamma(r_\tau) \la \sigma$ in a hot outflow with $L_\gamma \sim L_{\rm P}$.  
Radiation plays a key role in accelerating the outflow when this bound is satisfied.  

\subsection{Strong Radiative Acceleration Outside the Fast Critical Point}

An essential feature of a {\it steady} magnetohydrodynamic (MHD) flow
is that the equation for the fluid speed, obtained by combining the continuity and momentum equations,
becomes singular where it matches a normal mode of the fluid.  In the absence of significant matter
pressure, the relevant normal modes are the Alfv\'en mode and the fast mode.  

In a relativistic outflow, the inertia of the advected toroidal magnetic field rapidly
becomes insignificant outside the fast critical point.  Here we consider the case where
the poloidal magnetic flux surfaces do not experience the differential bending needed
for efficient MHD acceleration.  Then, when the radiative force is
strong enough to provide significant acceleration at the fast point, it also controls
the terminal acceleration of the outflow, as if the magnetic field were not present.
In spite of this, the Poynting flux carried by the magnetic field
can continue to dominate the kinetic energy flux:  the matter acts as a couple between
the radiation field and the magnetic field \citep{thompson06}.

To see this, focus on the zone far outside the Alfv\'en critical point.   The solution
to the induction equation is \citep{ferraro37}
\be\label{eq:induction}
\frac{B_{\phi}}{B_{r}} =
\frac{v_{\phi} - \Omega_f r\sin\theta}{v_{r}},
\ee 
where $\Omega_f$ is the rotational angular frequency of the magnetic footpoint
at the boundary of the engine, and is constant along a magnetic flux surface.
Equation (\ref{eq:induction}) implies a uniform rate of transport of toroidal magnetic flux
\be
{\dot\Phi}_\phi = v_r r B_\phi \simeq -\Omega \sin\theta r^2 B_r = {\rm const},
\ee
since the toroidal velocity has decreased to $v_\phi \ll v_r \simeq c$.  We now allow
an external (e.g. radiation) force to be applied to the fluid while imposing this
constraint, as well as the constancy of mass flux
\be
{d\dot M\over d\Omega} = \Gamma\rho v_r r^2 = {\rm const}.
\ee
The change in the radial energy flux is, at a fixed radius,
\be
\delta T_{tr} = \delta\left(\Gamma^2 \rho c^2 v_r + {E_\theta B_\phi\over 4\pi} c\right)
= {1\over r^2}\delta\left({d\dot M\over d\Omega}c^2 - {\dot\Phi}_\phi^2{1\over v_r}\right)
\simeq {\delta\Gamma\over r^2}\left({d\dot M\over d\Omega}c^2 - {{\dot\Phi}_\phi^2\over \Gamma^3 c}\right).
\ee
The right-hand-side here is singular at the fast point, where
\be
\Gamma^3 =  \Gamma_c^3 = {{\dot\Phi}_\phi^2\over c^3 d\dot M/d\Omega} = \sigma.
\ee
The (toroidal) magnetic field dominates the inertia of the outflow in between the Alfv\'en
and fast points, but the small flux of ordinary matter dominates outside the fast point.

Our main focus is on flows that are sub-fast magnetosonic at the transparency surface,
and are accelerated through the critical point by the radiation force as well as the Lorentz force.
For our adopted field geometry, this occurs when $\Gamma(r_\tau) \la \sigma^{1/3}$, and requires a strongly 
magnetized outflow.  

A second type of flow is super-fast magnetosonic at the transparency surface, 
having already been accelerated through the critical point, $\Gamma(r_\tau) \ga \sigma^{1/3}$.
The dominant acceleration mechanism outside the photosphere then depends on the poloidal field geometry.
When the poloidal flux surfaces do not diverge from each other, the case studied in this paper, the
Lorentz force freezes out and the remaining acceleration is by the radiation force.  However, in a jet
geometry there is a much more extended competition between MHD and radiation stresses outside the
fast critical point, as we discuss in paper II.

\subsection{Insensitivity to Flow Structure Near the Alfv\'en Critical Point}\label{s:alfvenc}

Since the Alfv\'en mode has the lower speed, its critical point sits closer to the rotating `engine'.  Inside
the Alfv\'en critical point, the magnetic field lines are effectively rigid and guide the outflow of matter and radiation.
Outside, the rotation of the engine 
causes the magnetic field lines to be bent back into a Parker spiral, and the magnetic field is mainly toroidal.  
After a choice is made for the poloidal field profile, we wish to test the {\it insensitivity} of the large-distance flow 
solution to details close to the Alfv\'en point.  In fact, we obtain an essentially unique outer solution without explicitly 
requiring it to pass through the Alfv\'en point.  If the solution were not unique in this way, it would be suspect:  
the magnetic field configuration near the Alfv\'en point maps directly onto the engine, and details of the inner
flow solution depend, in turn, on details of plasma heating close to the engine. 

\subsection{Bulk Acceleration by Magnetic Reconnection:  A Critique}

The acceleration of a cold, spherical MHD outflow is inefficient due to a cancellation between 
the magnetic pressure gradient and curvature forces.  Rather than considering deviations
from radial flow, \cite{drenkhahn02} suggest that reconnection of a reversing toroidal magnetic field
would convert Poynting flux to a large-scale relativistic expansion.  Inside the scattering
photosphere, they impose conservation of energy flux per steradian,
\be
{dL\over d\Omega} =  r^2\Gamma^2v_r \left[h + {(B_\phi/\Gamma)^2\over 4\pi}\right].
\ee
Here $h \simeq \rho c^2$ is the bulk-frame enthalpy per unit volume, $\rho$ is the proper 
rest-mass density, and $B_\phi/\Gamma$ the bulk-frame toroidal magnetic field.  A sink term for 
the magnetic flux is imposed, $c\partial_r (rB_\phi v_r) = -rB_\phi v_r/\tau$, but no corresponding 
source of $h$.  With these assumptions, the dissipation of Poynting flux must be compensated 
by a growth in kinetic energy.  A small decrease in Poynting flux corresponds to a large increase
in $\Gamma$ in a cold, high-$\sigma$ outflow.  

More realistically, reconnection heats the particles, and in a high-$\sigma$ 
outflow there is a large increase in the particle inertia.  Even allowing for 
rapid radiative cooling, the photons are trapped by the outflow, and
\be
\delta h \sim - \delta\left[{(B_\phi/\Gamma)^2\over 8\pi}\right].
\ee
Reconnection also creates a radial magnetic field, through the appearance
of multiple magnetic X-points.  The inertia of this small-scale magnetic 
field also dominates the particle rest mass.  Taking both of these effects 
into account, one sees that dissipation of even half the magnetic energy 
can generate only mildly relativistic bulk motion.

\section{Outflow Equations}\label{s:eom}

Given the technical challenge involved in adding the radiation force,
we are forced to make simplifying assumptions about the angular distribution of the poloidal magnetic flux.  In this first 
paper we follow \cite{goldreich70} in allowing the magnetic field to bend in the toroidal but not the poloidal directions --
i.e., the poloidal field is purely radial.  This approximation is, in fact, increasingly well justified 
for very intense radiation fields, e.g., those which are strong enough to force open magnetic field lines
inside the speed of light cylinder.  It should also hold with reasonable
accuracy close to the magnetic equator.  The solutions are then labeled by the fixed rest mass and radiation energy fluxes.

To the steady Euler equation for a cold, relativistic MHD fluid, we add a term representing the radiation force,
\be\label{e:Euler}
\rho\,\Gamma{\bf v}\cdot\bnabla(\Gamma{\bf v}) = {1\over 4\pi}\left[(\bnabla\cdot{\bf E}) {\bf E} + 
(\bnabla\times{\bf B})\times{\bf B}\right] + \frac{\Gamma\rho}{\bar{m}}{\bf F}^{\rm rad}.
\ee
Taking the dot product of equation 
(\ref{e:Euler}) with the poloidal magnetic field ${\bf B}_{\rm p}$ gives 
\ba\label{e:Euler1}
\left(B_{r}\partial_{r} + \frac{B_{\theta}}{r}\partial_{\theta}\right)\Gamma c^{2} 
&-& \frac{v_{\phi}}{r}\left[B_{r}\partial_{r}(r\Gamma v_{\phi}) 
+ \frac{B_{\theta}}{\sin\theta}\partial_{\theta}(\sin\theta\Gamma v_{\phi})\right] = \nn
&&\quad\quad\quad - \frac{B_{\phi}}{4\pi\Gamma\rho r}\left[B_{r}\partial_{r}(rB_{\phi}) 
+ \frac{B_{\theta}}{\sin\theta}\partial_{\theta}(\sin\theta B_{\phi})\right] 
+ \frac{1}{\bar{m}}\left(B_{r}F^{\rm rad}_{r}+B_{\theta}F^{\rm rad}_{\theta}\right).\nn
\ea 
The projection of the Coulomb force onto ${\bf B}_p$ vanishes in a steady, axisymmetric MHD
outflow with $E_\phi = 0$, since then ${\bf E}\cdot{\bf B} = {\bf E}_p \cdot {\bf B}_p$ = 0.
The $\phi$-component of equation (\ref{e:Euler}) is
\be\label{e:Euler2}
\frac{v_{r}}{r}\partial_{r}(r\Gamma v_{\phi}) 
+ \frac{v_{\theta}}{r\sin\theta}\partial_{\theta}(\Gamma v_{\phi}\sin\theta) 
= \frac{1}{4\pi\Gamma\rho r}\left[B_{r}\partial_{r}(rB_{\phi}) 
+ \frac{B_{\theta}}{\sin\theta}\partial_{\theta}(\sin\theta B_{\phi})\right] 
+ \frac{1}{\bar{m}}F^{\rm rad}_{\phi}.
\ee 
The toroidal and poloidal components of ${\bf B}$ are related through the flux-freezing condition (\ref{eq:induction}).

The photon emission radius $r_s$ serves as a reference length, and the photon compactness
is also measured at this radius:
\be
x\equiv {r\over r_s}; \quad\quad  \omega \equiv {\Omega_f r_s\over c};\quad\quad \chi_s\equiv \frac{\sigma_{T}L_\gamma}{4\pi r_s\bar{m}c^{3}}.
\ee  
In a GRB outflow, the photosphere generally lies outside the light cylinder
of the rotating engine, so we take $\omega = \Omega_f r_s/c > 1$ in our calculations.  For example, 
an outflow launched by a millisecond engine which fills an opening angle $\theta \sim 0.1$ at a breakout
radius of $\sim 10^{10}$ cm will have a physical width $\Omega_f r \sin\theta/c \sim 100$ times larger
than the engine light cylinder.\footnote{Early claims of measurements of jet opening angles based on
temporal breaks in GRB afterglow light curve (e.g. \citealt{rhoads99}) have been revealed to be somewhat
ambiguous based on more complicated behavior seen in the Swift data (e.g. \citealt{zhang06}).  
Nonetheless in a few cases detailed fits with direct hydrodynamical modelling are consistent with
opening angles $\sim 0.1$ (e.g. \citealt{vaneerten12}), and a range of a few around this value seems
likely (e.g. \citealt{pk02}).}  This means that only a tiny fraction $\sim 10^{-5}-10^{-3}$ of the
solid angle of a laminar jet will rotate rapidly enough that our approximation breaks down.  (In practice, a
modest amount of turbulence in the jet could mix this thin inner cone with the much wider, slowly rotating
annulus surrounding it.)  In practice, for computational ease, we consider intermediate values of $\omega$.
In the case of an isolated millisecond magnetar one might indeed have $\omega\sim1$.

The magnetization is defined by
\be\label{eq:sigma}
\sigma \equiv 
\frac{B_r^2\Omega_f^2r^2\sin^2\theta}{4\pi \Gamma\rho v_rc^3}
\ee 
which is constant in the monopolar outflow considered here.\footnote{This definition, following \cite{michel69} and \cite{goldreich70}, differs in terms $O(v_\phi/\Omega_f r\sin\theta_f)$ to both 
$\sigma = (c^2d\dot M/d\Omega)^{-1}dL_P/d\Omega$, and
$\sigma = (c^3 d\dot M/d\Omega)^{-1}\dot \Phi_\phi^2$, where 
$\dot\Phi_\phi = v_r r \sin\theta_f B_\phi$ is the advection rate of toroidal flux. With this definition $\sigma$ is independent of radius if the poloidal field is restricted to be purely radial, 
but the last two definitions are non-constant at $O(v_\phi/\Omega_f r\sin\theta_f)$.}  
The kinetic, Poynting and radiation luminosities, $\Gamma\dot Mc^2$, $L_{\rm P}$ and $L_\gamma$, are related by 
\be\label{eq:lmax}
\frac{L_{\rm P}}{\Gamma\dot Mc^2} =
-\frac{\sigma}{\Gamma x\omega\sin\theta}\frac{B_{\phi}}{B_{r}} \sim \frac{\sigma}{\Gamma};
\quad\quad
\frac{L_\gamma}{\Gamma\dot Mc^2} \sim {1\over 6\Gamma^2(x_\tau)x_\tau}{\chi_s\over \Gamma},
\ee 
where in the last equality we have have assumed that the flow is relativistic [see equation (\ref{eq:rtau})]. 
When radiation is absent, the energy and angular momentum per unit rest mass,
\be\label{eq:integral}
\mu =
\Gamma - \frac{\sigma}{x\omega\sin\theta}\frac{B_{\phi}}{B_{r}}; \quad\quad 
\mathcal{L} =
  \Gamma x \sin\theta \beta_{\phi} - \frac{B_{\phi}\sigma}{B_r\omega^2x\sin\theta}
\ee 
are conserved along field lines.

\subsection{Relativistic Radiation Force}\label{s:rad_force}

The radiation emanates from a spherical static `emission surface' of radius $r_s$ ($x = 1$).  We
adopt simplified spectral and angular distributions:  the unscattered radiation is monochromatic,
uniform at angles $0 < \theta < \pi/2$ at the emission surface, 
and streams freely outward.  At $x>1$, the cone of the radiation field contracts and
\be\label{eq:thetas}
I_\nu = I_0\nu_0\delta(\nu-\nu_0) \quad{\rm for}\quad \theta < \theta_s \equiv 
\sin^{-1}\left({r_s\over r}\right) = \sin^{-1}\left({1\over x}\right);
\quad\quad
I_\nu = 0\quad{\rm for}\quad \theta > \theta_s.
\ee
This accurately represents an isotropically 
emitting star that is surrounded by an optically thin wind. 
It still produces qualitatively correct results if the outflow is optically thick near the 
engine, and experiences nearly linear growth of Lorentz factor with radius, 
driven by radiation pressure (Section \ref{s:relphot}).

Because $\omega>>1$ typically, a non-rotating emission surface is a reasonable
approximation.  The effect of adding modest rotation to the photon source is addressed quantitatively
in Appendix \ref{s:rotation}, and the corrections to the flow solution are shown to be small.

The radiation interacts with electrons (and positrons) via Thomson scattering, and the radiative force
is taken to be unperturbed by this interaction.  The radiation force per scattering charge 
in a global inertial (`lab') frame is related to the force ${\bf F}^{\rm rad\,'}$ in the flow rest frame by
\be\label{e:Flab}
{\bf F}^{\rm rad} =
\left[{\bf F}^{\rm rad\,\prime} 
- \left({\bf F}^{\rm rad\,\prime}\cdot\bbeta\right)\frac{\bbeta}{\beta^{2}}\right]\Gamma^{-1} 
+ \left({\bf F}^{\rm rad\,\prime}\cdot\bbeta\right)\frac{\bbeta}{\beta^{2}},
\ee 
where primes denote quantities in the frame co-moving with the fluid. 
Letting $\hat{k}$ be the unit wave vector of the incoming photon, we have the transformations
\ba\label{e:trans}
I_{\nu}^{\prime} =
I_{\nu}\Gamma^{3}(1-\bbeta\cdot\hat{k})^{3}; \quad\quad
d\nu^{\prime} &=&
d\nu\Gamma(1-\bbeta\cdot\hat{k}); \quad\quad
d\Omega^{\prime} =
\frac{d\Omega}{\Gamma^{2}(1-\bbeta\cdot\hat{k})^{2}};\nn
\hat{k}^{\prime} &=&
\frac{\hat{k} + (\Gamma-1)(\hat{\beta}\cdot\hat{k})\hat{\beta} - \bbeta\Gamma}{\Gamma(1-\bbeta\cdot\hat{k})}.
\ea
The radiation force in the fluid frame is given by 
\be
{\bf F}^{\rm rad\,\prime}\equiv
\frac{d{\bf p}^{\prime}}{dt^{\prime}} 
= \frac{\sigma_{T}}{c}\int I_{\nu}^{\prime}\hat{k}^{\prime}d\Omega^{\prime}d\nu^\prime 
= \frac{\sigma_{T}I_{0}\nu_0}{c}\int\Gamma^{2}\left(1-\bbeta\cdot\hat{k}\right)\left[\hat{k}\Gamma^{-1}
+ (1-\Gamma^{-1})\left(\frac{\bbeta\cdot\hat{k}}{\beta^{2}}\right)\bbeta-\bbeta\right]d\Omega,
\ee 
and combining with equation (\ref{e:Flab}) gives 
\be\label{Flab2}
{\bf F}^{\rm rad} =
\frac{\sigma_{T}I_{0}\nu_0}{c}\int\left(1-\bbeta\cdot\hat{k}\right)\left[\hat{k}
-\bbeta\Gamma^{2}\left(1 - \bbeta\cdot\hat{k}\right)\right]d\Omega.
\ee 

The radiation force (\ref{Flab2}) can be evaluated analytically by integrating over the emission surface.
Defining dimensionless functions $R$, $P$, by
\be\label{eq:fdef}
F^{\rm rad}_r = \chi_s {\bar{m} c^2\over r_s} R(r,\Gamma);\quad  F^{\rm rad}_\phi = \chi_s {\bar{m} c^2\over r_s} P(r,\Gamma),
\ee
one finds
\be\label{e:R}
R = -\frac{8}{3}u\Gamma + \frac{1+2u^{2}}{x^{2}} 
+ \beta_{r}\left(2\Gamma^{2}+v^{2}\right)\sqrt{1-\frac{1}{x^{2}}} 
+ \beta_{r}\left(\frac{2}{3}\Gamma^{2} - v^{2}\right)\left(1-\frac{1}{x^{2}}\right)^{3/2};
\ee 
\be\label{e:P}
P = -\beta_{\phi}\left[\frac{8}{3}\Gamma^{2}-\frac{2u\Gamma}{x^{2}} 
- \left(1+2\Gamma^{2}+v^{2}\right)\sqrt{1-\frac{1}{x^{2}}}
+\frac{1}{3}\left(1-2u^{2}+v^{2}\right)\left(1-\frac{1}{x^{2}}\right)^{3/2}\right],
\ee 
where the radial and non-radial four-velocities are
\be
(u,v)\equiv
(\Gamma\beta_r,\Gamma\beta_{\phi}).
\ee 

At large radius and Lorentz factor, expressions (\ref{e:R}) and (\ref{e:P}) simplify to
\be
R\simeq
\frac{1}{4x^2\Gamma^2} - \frac{\Gamma^2}{12x^6}; \quad\quad
P\simeq
-\frac{\beta_{\phi}}{2x^2}\left(\frac{1}{2\Gamma^2} 
+ \frac{1}{x^2} + \frac{\Gamma^2}{6x^4}\right).
\ee 
The first term in the radial force is due to photons propagating nearly parallel to the fluid;
the second is the drag caused by photons which are aberrated into the anti-radial
direction by the relativistic particle motion.  The photon field seen by the particle is nearly
isotropic in a frame were $R = 0$.  At large $\chi_s$ and $x$ not too large, 
the matter therefore tends toward the equilibrium Lorentz factor
\be\label{eq:gameq}
\Gamma_{\rm eq}\simeq 3^{1/4}\left({r\over r_s}\right) = 3^{1/4}x.
\ee 
This frame only exist due to the extended nature of the source.

\subsection{Photon Distribution at a Displaced Photosphere}\label{s:relphot}

In many cases, the outflow moves relativistically at its scattering photosphere, so
that the radiation field is already collimated at the base of the transparent zone.
Then the emission surface becomes a virtual one.   Setting the photospheric Lorentz factor
to the equilibrium value (\ref{eq:gameq}), the emission surface is pushed inward to a radius 
\be\label{eq:rseff}
r_{s,\rm eff} = 3^{1/4}\,{r_\tau\over \Gamma(r_\tau)}.
\ee

A radiation field that emerges from a relativistically moving
photosphere is not cut off sharply at an angle $\theta_s = r_{s,\rm eff}/r_\tau = 3^{1/4}/\Gamma(r_\tau)$,
but has a somewhat smoother cutoff.   Consider, instead, an intensity that is isotropic in the 
matter rest frame, $I'_{\nu'} = I_0^R\nu_0^R \delta(\nu'-\nu_0^R)$, and compare
with the top-hat spectral distribution (\ref{eq:thetas}) by normalizing to a fixed flux in
the frame of the engine, $F = 2\pi\int d\mu \int d\nu I_\nu(\mu)$.  Then the frequency-integrated
intensity in the lab frame is
\be 
I(\theta) = \int d\nu I_\nu(\theta) = {I_0^R\nu_0^R\over [\Gamma(1-\beta\mu)]^4}  \simeq 
{I(\theta=0)\over (1 + \Gamma^2\theta^2)^4} 
= {3^{3/2}\over (1 + \sqrt{3}\theta^2/\theta_s^2)^4} {F\over \pi\theta_s^2}
\quad\quad\left(\Gamma = \Gamma_{\rm eq} = {3^{1/4}\over\theta_s}\right),
\ee
as compared with $I = F/\pi \theta_s^2$ for $\theta < \theta_s$.  

These two angular distributions yield the same photon force for a static charge at $r \gg r_{s,\rm eff}$ 
and, by construction, the same Lorentz factor for which the radiation force vanishes.
The power with which an electron scatters photons in the bulk frame is nearly
identical as well:  $P' = 4\pi \int d\nu' \sigma_T I'_{\nu'} = 3\sigma_TF/4\Gamma^2$ for the isotropic
bulk-frame intensity, and 
\be
P' = \left(1 + \sqrt{3}{\Gamma^2\over\Gamma_{\rm eq}^2} + {\Gamma^4\over\Gamma_{\rm eq}^4}\right){\sigma_TF\over 4\Gamma^2}
= \left({1\over 2} + {\sqrt{3}\over 4}\right){\sigma_TF\over\Gamma_{\rm eq}^2} \quad\quad (\Gamma = \Gamma_{\rm eq})
\ee
for the top-hat distribution.  
The spectrum of photons scattered by a cold electron flow is proportional to $P'$, and differs in normalization by $\sim 20\%$ in the two cases.

More generally, if the outflow interacts with slower relativistic material -- as discussed by \cite{thompson06} in the
context of GRBs -- then the beam of photospheric photons is scattered into a cone of width $> 1/\Gamma$.  This motivates our
considering outflows with i) significant scattering optical depth at $r = r_s$; but ii) $\Gamma(r_s) > 1$. 
Given the potential importance of such additional effects for GRBs, we stick with the simplest model
a photon emission surface with uniform intensity, with the understanding that this a virtual surface in some circumstances.
See Section \ref{s:spectrum} for calculations of the scattered photon spectrum and the application to GRBs.

\begin{figure}[h]
\centerline{\includegraphics[width=0.5\hsize]{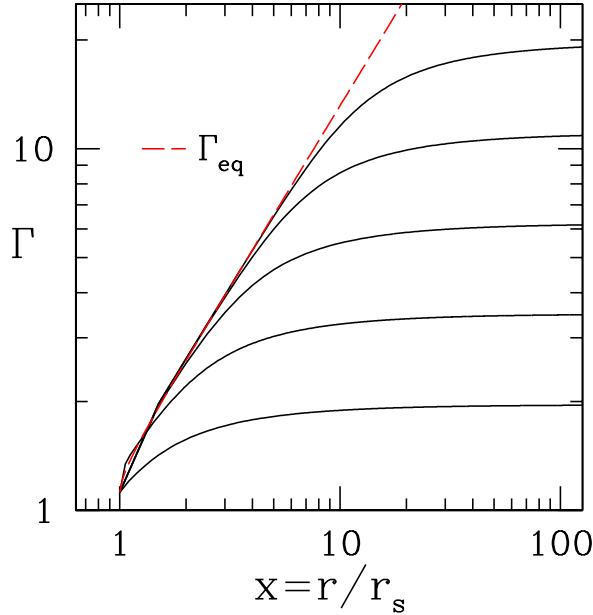}}
\caption{Radiative acceleration of an unmagnetized outflow by a photon source at $x =1$.  
The compactness $\chi_s$ is varied from $10-10^5$.  Here $\Gamma_{\rm eq}=3^{1/4}x$.}
\label{fig:unmagnetized}
\vskip .4in
\end{figure}

\subsection{Acceleration of an Unmagnetized Outflow}\label{s:unmag}

As a test of our formalism, we revisit the acceleration of cold matter by an intense radiation field,
setting $\sigma = 0$.  Then equations (\ref{e:Euler}), (\ref{eq:fdef}) and (\ref{e:R}) combine to give
\be
{d\Gamma\over dx} = 
\chi_s\left(R + \frac{v}{u}P\right); \quad\quad
{d\beta_{\phi}\over dx} = 
-\frac{\beta_{\phi}}{x} - \chi_s\left[\frac{\beta_{\phi}}{\Gamma}R 
- \frac{(1-\beta_{\phi}^{2})}{u}P\right], 
\ee 
and simplify further to 
\be
{d\Gamma\over dx} \simeq
\chi_s R\simeq
\frac{\chi_s}{4x^2}\left(\frac{1}{\Gamma^2} - \frac{\Gamma^2}{3x^4}\right); \quad\quad
{d\beta_{\phi}\over dx} \simeq
-\frac{\beta_{\phi}}{x} 
- \frac{\chi_s\beta_{\phi}}{2x^2\Gamma}\left(\frac{1}{\Gamma^2}+\frac{1}{x^2}\right)
\ee 
for large $x$, $\Gamma$.  The matter Lorentz factor tracks $\Gamma_{\rm eq}$ until the local compactness 
$\chi \sim \chi_s/\Gamma^3 x$ drops below $\sim \Gamma$ (see Figure \ref{fig:unmagnetized}).  The asymptotic Lorentz
factor is
\be
\Gamma_{\infty}\simeq
1.086\chi_s^{1/4}
\ee 
as long as baryon loading is sufficiently low ($\eta\equiv L_\gamma/\dot Mc^2 \gg \chi_s^{1/4}$); 
otherwise $\Gamma_{\infty}\simeq \eta$.

The radiation force has a stronger effect when
the flow remains optically thick out to a large distance from the engine, and
develops at least a moderate Lorentz factor before the photons begin to stream freely.  

In this situation, the constraint $L_\gamma < L_{\rm P}$ would correspond to a limit on the photon compactness as
measured at the photosphere,
\be
\chi(r_\tau) \equiv {L_\gamma\sigma_T\over 4\pi r_\tau m_ec^3} = {r_s\over r_\tau} \chi_s,
\ee
namely,
\be
\chi(r_\tau) \la 6\Gamma^2(r_\tau) \sigma
\ee
[see equation (\ref{eq:lmax})].  Now the parameter $\chi_s$ describes a `virtual' photon source
that is buried at large scattering depth, and is already collimated at the photosphere.
The terminal Lorentz factor increases to
\be
\Gamma_{\infty,\chi} \;\sim\; \chi_s^{1/4} = \left[3^{-1/4}\,\Gamma(r_\tau)\,\chi(r_\tau)\right]^{1/4}.
\ee
If the fast point were to sit in the transparent zone, then $\Gamma(r_\tau) \la
\sigma^{1/3}$ and one has the bound
\be\label{eq:chitau2}
\chi_\tau \la 6\sigma^{5/3};\quad\quad \Gamma_{\infty,\chi} \la 1.3\sigma^{1/2}.
\ee

If the bound (\ref{eq:chitau2}) is not satisfied, then the outflow will pass through the
fast critical point before reaching its photosphere.  Nonetheless, the radiation field will 
provide significant supplemental acceleration outside the photosphere for any value
of $\Gamma(r_\tau)$ below $\sigma$.

\section{Dimensionless Outflow Equations:  $\theta \la \pi/2$}\label{s:mono}

The poloidal magnetic field lines make a transition from dipolar to nearly radial at the light cylinder of an isolated pulsar,
being forced open by causality constraints \citep{contopolous05, mckinney06, spitkovsky06}.   More collimated outflows are expected 
from magnetized stars surrounded by accretion disks \citep{bz77,bp82}, and are the focus of paper II.

We consider a streamline close to the magnetic equator (but not so close as to involve the equatorial current sheet).  
We work with a simplified form of the Euler equations that includes the radiation force, but assumes the poloidal 
magnetic field to be purely radial. Therefore we substitute
\be
B_r \simeq \frac{B_{r,s}}{x^{2}}; \quad\quad \frac{B_{\theta}}{B_r} = \frac{v_{\theta}}{v_r} \ll 1; \quad\quad\sin\theta \simeq 1,
\ee 
into equations (\ref{e:Euler1}), (\ref{e:Euler2}), which gives
\be\label{eqn:Eulerapp1}
\partial_{r}\Gamma c^{2} - \frac{v_{\phi}}{r}\partial_{r}(r\Gamma v_{\phi}) =
-\frac{B_{\phi}}{4\pi\Gamma\rho r}\partial_{r}(rB_{\phi}) 
+ \frac{1}{\bar{m}}F^{\rm rad}_{r};
\ee 
\be\label{eqn:Eulerapp2}
\frac{v_{r}}{r}\partial_{r}(r\Gamma v_{\phi})=
\frac{B_{r}}{4\pi\Gamma\rho r}\partial_{r}(rB_{\phi}) 
+ \frac{1}{\bar{m}}F^{\rm rad}_{\phi}.
\ee 
The various terms in the right-hand sides of these equations can be separated into purely magnetocentrifugal pieces (which do not
depend on the radiation force), the direct radiation force, and a cross term:
\be\label{eqn:Gamp}
{d\Gamma\over dx} = 
\frac{\Gamma'_{\sigma} + \Gamma'_{\chi} + \Gamma'_{\sigma\chi}}{\mu_{\rm eff}}; \quad\quad
{d\beta_{\phi}\over dx} =
\frac{\beta'_{\phi,\sigma} + \beta'_{\phi,\chi} + \beta'_{\phi,\sigma\chi}}{\mu_{\rm eff}},
\ee 
where
 \be\label{eqn:gammaprime}
\Gamma^{\prime}_{\sigma} + \Gamma^{\prime}_{\chi} + \Gamma^{\prime}_{\sigma\chi} \;\equiv\;
-\frac{\sigma}{\beta_{r}x\omega}\left[\frac{\beta_{\phi}}{x}\left(2+\frac{\beta_{\phi}}{\beta_{r}}\frac{B_{\phi}}{B_{r}}\right)\right] 
+ \chi_s\left(R+\frac{\beta_{\phi}}{\beta_{r}}P\right)
- \frac{\sigma\chi_s}{ux^{2}\omega^{2}}\left(1+\frac{\beta_{\phi}B_{\phi}}{\beta_{r}B_{r}}\right)\left(R+\frac{B_{\phi}}{B_{r}}P\right);
\ee 
\begin{eqnarray}\label{eq:betaprime}
\beta^{\prime}_{\phi,\sigma} + \beta^{\prime}_{\phi,\chi} + \beta^{\prime}_{\phi,\sigma\chi} &\;\equiv\;&
-\frac{\beta_{\phi}}{x}\left[1+\frac{\sigma}{ux^{2}\omega^{2}} -\frac{\sigma}{u^{3}} 
 - \frac{\beta_{\phi}\sigma}{ux\omega}\left(1-\frac{1}{u^{2}}\right)\right] \nn
&&\quad -\,\chi_s\left[\frac{\beta_{\phi}}{\Gamma}R-\frac{(1-\beta_{\phi}^{2})}{u}P\right]-
\frac{\sigma\chi_s}{x^{2}\omega^{2}\Gamma^{2}u^{2}}\frac{B_{\phi}}{B_{r}}\left(R+\frac{B_{\phi}}{B_{r}}P\right),
\end{eqnarray}
and
\be\label{eq:mueff}
\mu_{\rm eff} =
1 - \frac{\sigma}{u^{3}}\left(1+v^{2}\right) 
- \frac{\sigma}{ux^{2}\omega^{2}}\left(1+\frac{v^{2}}{u^{2}}\right) 
+ \frac{2\sigma v\Gamma}{u^{3}x\omega}.
\ee 

\subsection{Flow Through the Fast Critical Point:  $\theta\la \pi/2$, $B_r \sim r^{-2}$}\label{s:fcrit}
These MHD wind equations have critical points where the flow matches the speed of
a cold MHD mode.   The Alfv\'en critical point corresponds to a flow speed $u=\sigma(1-\omega^2x^2)/\omega^2x^2$, 
and therefore sits inside the light cylinder.  

Since we are looking for robust flow solutions which do not depend on the magnetic field structure 
in this inner zone (Section \ref{s:alfvenc}), our focus is on the fast critical point, where
\be\label{e:msspeed}
u = \frac{\sigma }{x^2\omega^2}\left(1-\omega^2x^2+\frac{B_{\phi}^2}{B_r^2}\right).
\ee 
This is the singularity appearing in equation (\ref{eqn:Gamp}).  A regular flow solution passing 
through this point must satisfy two conditions:
\be\label{eq:regularity}
\Gamma'_{\sigma}(x_c) + \Gamma'_{\chi}(x_c) + \Gamma'_{\sigma\chi}(x_c) = 0; \quad\quad \mu_{\rm eff}(x_c) = 0.
\ee  
Here $x_c$ is the (so-far undetermined) radius of the fast point.
Equations (\ref{eq:regularity}) generate a one parameter family of flow solutions.  

The specific angular momentum evolves according to the simple equation $d\mathcal{L}/dx = (x/\beta_r)\chi_s P$.
We have tested equations (\ref{eqn:Eulerapp1}) and (\ref{eqn:Eulerapp2}) by combining them and re-deriving this
equation.  This means that the critical point of the equation for $d\beta_\phi/dx$ does not impose independent 
constraints on the flow solution:  a solution with a smooth $\Gamma(x)$ profile automatically satisfies the 
$\beta'_{\phi,\sigma}(x_c) + \beta'_{\phi,\chi}(x_c) + \beta'_{\phi,\sigma\chi}(x_c) = 0$.

The presence of a cross term in the wind equations, involving both the magnetization $\sigma$ and the compactness $\chi_s$, 
deserves some comment.  Inside the fast magnetosonic point, and outside the Alfv\'en point, 
the {\it radial} motion of the fluid has effectively a negative inertia:  a positive external radial 
force extracts energy from the outflow.  This negative inertia arises from the response of 
$B_\phi$, which dominates the energy integral (\ref{eq:integral}), to changes 
in the radial flow speed.   The scaling $B_\phi \sim \beta_r^{-1}$ implies a decrease in toroidal
field energy with increasing $\beta_r$.  On the other hand, the angular momentum 
(\ref{eq:integral}) is also dominated by the electromagnetic field.  A change in $B_\phi$ creates
unbalanced toroidal stresses, which are a source for $\beta_\phi$, and are proportional to
$\sigma\chi_s$.   Changes in $\beta_\phi$ and $\beta_r$ of opposing signs
allow $B_\phi$ to remain nearly constant.  The net change in Lorentz factor is positive,
$\delta \Gamma = \Gamma^{-3}(\beta_r\delta\beta_r + \beta_\phi\delta\beta_\phi) > 0$, if
$1 + \beta_\phi B_\phi/\beta_r B_r > 0$.   Close to the light cylinder, where the term
$\sim -\sigma/u x^2\omega^2$ in $\mu_{\rm eff}$ dominates the term $-\sigma/u^3$, the effective inertia has
the usual sign.

\subsection{Pure MHD Wind ($\chi_s=0$) with Monopolar Radial Magnetic Field}
When radiation fields are absent and $\Gamma \gg 1$ our equations reduce to the cold limit of the 
system studied by \cite{goldreich70}. 
The angular momentum ${\cal L}$ [equation (\ref{eq:integral})] is conserved, and determines
 \be
\beta_{\phi} =
\frac{x\omega(\mathcal{L}\omega\beta_r - \sigma)}{x^{2}\omega^{2}u - \sigma}.
\ee 
The fast point lies at infinite radius, the Lorentz factor being limited to 
its critical value 
\be
\Gamma_{\infty,\sigma} = \sqrt{1 + \sigma^{2/3}} \simeq \sigma^{1/3}.
\ee 
The unique solution passing through the Alfv\'{e}n and fast critical points is the
minimum-energy solution found by \cite{michel69}.  The slow acceleration of radial flows 
is an artifact of the near perfect cancellation of the outward magnetic pressure gradient force and
the inward curvature force.  Faster acceleration is possible through a faster-than-spherical divergence 
of the outflow \citep{begelman94}, or differential bending of
the poloidal field lines \citep{tchek09}, the effect of which we examine in paper II.

\subsection{Small Compactness Limit: $\chi_s\ll 4\sigma^{4/3}\omega^{-2}$}
For small but finite $\chi_s$, the Lorentz factor at the fast point remains unchanged from the 
pure MHD solution, $\Gamma_{c} \simeq \sigma^{1/3}$,
but the critical point is brought in to a finite radius.
The pure magnetocentrifugal and radiation terms in the wind equations dominate at large $x,\Gamma$, and simplify to
\be
\Gamma^{\prime}_{\sigma} \simeq
-\frac{v\sigma}{ux^2\omega}\left(2 - \frac{\beta_{\phi}x\omega}{\beta_r^2}\right) \simeq 
-\frac{v\sigma}{ux^2\omega}; \quad\quad
\Gamma^{\prime}_{\chi} \simeq
\chi_s R\simeq\frac{\chi_s }{4x^2\Gamma^2}
\ee 
Applying the regularity condition $\Gamma^{\prime}_{\sigma}+\Gamma^{\prime}_{\chi}\simeq0$ at the critical point gives
\be\label{eq:lowcomp}
x_{c}\simeq
\frac{4\sigma^{5/3}}{\chi_s \omega^2}; \quad\quad
\beta_{\phi,c} \simeq
\frac{1}{x_c\omega} \simeq
\frac{\chi_s \omega}{4\sigma^{5/3}}; \quad\quad
\mathcal{L}_{c} \simeq \Gamma_cx_c\beta_{\phi,c} + \frac{\sigma}{\omega \beta_{r,c}} \simeq
{\sigma^{1/3}\over\omega}\left(1+\sigma^{2/3}\right)\quad\quad({\rm low}~\chi_s).
\ee

\subsection{Large Compactness Limit: $\chi_s\gg 4\sigma^{4/3}\omega^{-2}$}\label{s:largechi}

For very large $\chi_s$ the fluid is locked to the radiation field while 
crossing the fast point at a finite radius,
\be\label{eq:highchixc}
\Gamma_c \simeq \Gamma_{\rm eq} \quad \Rightarrow \quad x_c \simeq 3^{-1/4}\Gamma_c\quad\quad({\rm high}~\chi_s).
\ee 
Here $B_r$ need not be small compared with $B_\phi$.
Approximating $\beta_\phi \ll x\omega$ and making use of equation (\ref{eq:induction}), equation (\ref{e:msspeed}) becomes
\be 
u^{3}\simeq
\sigma\left(1+\Gamma^{2}\frac{B_{r}^{2}}{B_{\phi}^{2}}\right) \simeq
\frac{B_{\phi}^{2}+\Gamma^{2}B_{r}^{2}}{4\pi u\rho c^{2}}
\ee
Substituting $u_c\simeq\Gamma_c\simeq 3^{1/4}x_c$ gives the critical point parameters
\be\label{eq:highcomp}
\Gamma_{c}\simeq \sigma^{1/3}\left(1+\frac{\sqrt{3}}{\omega^2}\right)^{1/3}\quad\quad
\mathcal{L}_c \simeq \frac{\sigma}{\omega}\left(1 + \frac{1}{2}\Gamma_{c}^{-2}\right)
\quad\quad({\rm high}~\chi_s).
\ee  
The transition between the low- and high-$\chi_s$ scalings for the fast point
occurs where $x_{c,\rm low}\simeq x_{c,\rm high}$, at a compactness $\chi_s \simeq
4\sigma^{4/3}\omega^{-2}$.


\subsection{Numerical Methods and Boundary Values}\label{s:numericalmeth}
\begin{figure}[h]
\centerline{\includegraphics[width=0.6\hsize]{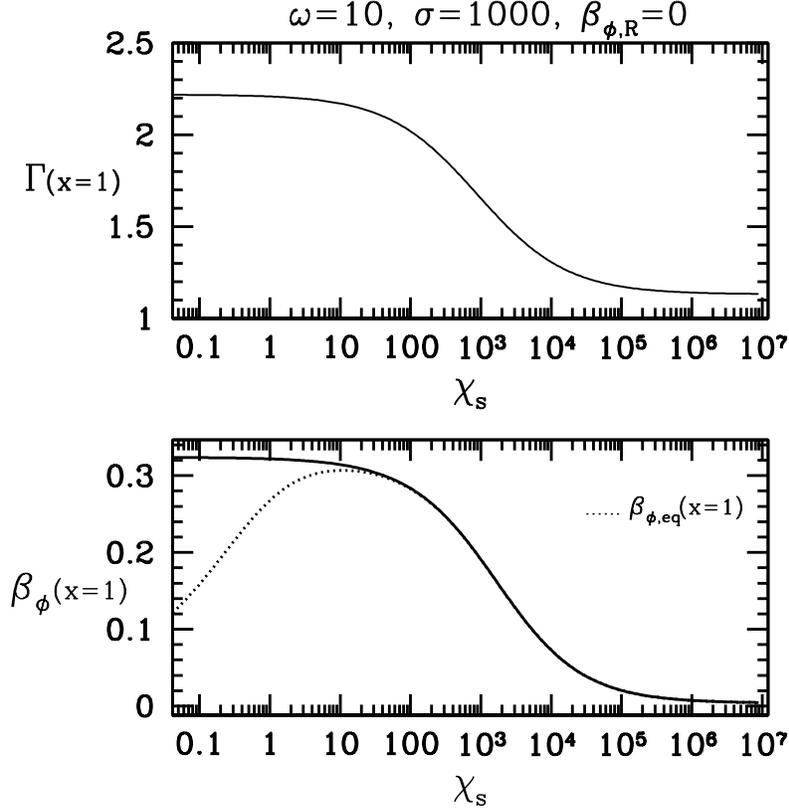}}
\caption{Lorentz factor and rotation speed at the inner boundary of an outflow
with $\sigma=10^3$, $\omega\equiv\Omega_fr_s/c=2$.
Dotted line corresponds to a vanishing net toroidal acceleration from the radiation field
at the inner boundary, equation (\ref{eq:betaphibc}).}
\label{fig:initialpoints}
\vskip .2in
\end{figure}

The singularity at the fast point is difficult to handle with standard integration methods,
especially when the wind equations become stiff, as they do when $\chi_s \gg 1$.  We employ two different 
integration schemes:  a shooting method for low $\chi_s$, and a relaxation method for high $\chi_s$.  
The position $x_c$ of the critical point is unknown {\it a priori}, since
the regularity condition provides only one constraint on the two coupled ODEs.
Once a candidate value of $x_c$ is chosen,
the flow variables at the critical point are uniquely determined, as is the flow solution interior to
to it.  However, this interior solution generally diverges at small radius.  A strong divergence is avoided
only for a narrow range of $x_c$, and even then the solution tends to develop sharp gradients in 
$\Gamma$ and $\beta_\phi$.  
Avoiding such gradients leads to an essentially unique choice of $x_c$ and a robust flow solution.
We have found that these smooth solutions have the property that the terms in $d\beta_\phi/dx$
which depend explicitly on the radiation field should sum to zero at $x = 1$,
\be\label{eq:betaphibc} 
\left(\beta^{\prime}_{\phi,\chi}+\beta^{\prime}_{\phi,\sigma\chi}\right)_{x=1}=0.
\ee
This recovers Michel's minimum-energy solution \citep{michel69} in the limit of vanishing radiation field, 
$\chi_s\rightarrow0$.  The resulting initial values of $\Gamma$ and $\beta_{\phi}$ are plotted in Figure 
\ref{fig:initialpoints} for $\sigma=10^3$ and $\omega=2$. 

At low $\chi_s$, the flow solution interior to $x_c$ is obtained by first determining the locus of critical points,
and the corresponding values of $\Gamma(x_c)$, $\beta_{\phi}(x_c)$.  We then shoot inward from $x=x_c$
using a 5th-order Runge-Kutta algorithm with adaptive step size (see Sections 7.3, 7.5 of \citealt{kiusalaas10}). 
The value of ${\cal L}_c$ -- which, unlike $x_c$, is single valued -- is iterated until the required small-$x$ behavior is obtained.  
This method fails when $\chi_s\gtrsim100$, since the equations become extremely stiff near $x=1$, and machine precision 
becomes inadequate to distinguish values of $\{x_c, \Gamma(x_c), \beta_\phi(x_c)\}$ that lead to converging and diverging solutions.

At high $\chi_s$, we use the relaxation method described in \cite{london82}
for transonic hydrodynamic flows.   The wind equations are replaced by finite-difference equations on a grid.  Starting
with a simple trial solution, and an initial guess for $x_c$, the inner boundary condition (\ref{eq:betaphibc}) is
applied along with the regularity condition (\ref{eq:regularity}) at the critical point.  
Since the location of the critical point is unknown, an independent variable $q$ which labels mesh points is introduced, 
along with a mesh-spacing function $Q(x)=\Psi q$ (where $\Psi$ is an unknown number).  We choose $Q(x)\propto\ln x$, which
tightly packs the grid points near $x=1$ (where the equations are stiff), and spreads them out near the critical point 
(so as to avoid divergences induced by the singularity).
This necessitates adding two more ODEs, for $x(q)$ and $\Psi$.  The error in the initial
guess at grid point $i$ is quantified in terms of $E_i = y_i - y_{i-1} - (x_i-x_{i-1})dy/dx$ 
and a new solution is obtained by a multi-dimensional Newton-Raphson method (see section 18.3
of \citealt{press07}).   The matrix inversion is performed using a Gaussian pivoting algorithm detailed in 
\cite{kiusalaas10}.  The critical point must be approached from 
below with this method and so the trial solution is typically cut off at $\sim 90 \%$ of the expected critical radius. 
This method fails in the low $\chi_s$ regime since the critical point is at large radius where $d\Gamma/dx \simeq 0$,
and even a slight overshoot in $\Gamma_c$ will be catastrophic. 

In both the low and high $\chi_s$ regimes, the solution is completed by shooting outward from the critical point.

\begin{figure}[h]
\centerline{\includegraphics[width=0.62\hsize]{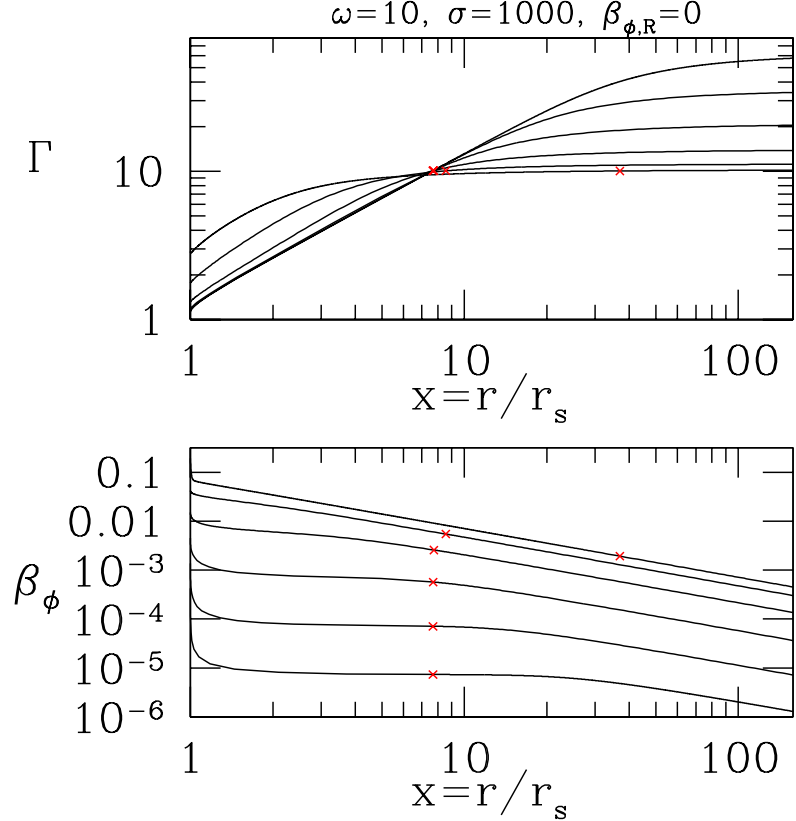}}
\caption{Acceleration of a magnetized outflow ($\sigma=10^3$, $\omega\equiv\Omega_fr_s/c=10$) for varying compactness
of a radiation source situated at $x = 1$:  $\chi_s=10^2-10^7$, bottom-top on the right side
in $\Gamma(x)$, top-bottom in $\beta_\phi(x)$.}
\label{fig:gammabeta}
\end{figure}
\begin{figure}[h]
\centerline{\includegraphics[width=0.55\hsize]{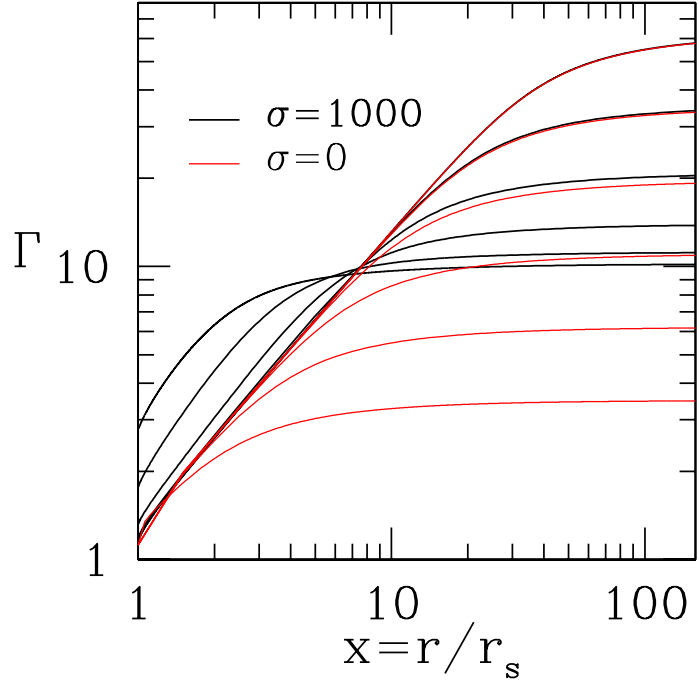}}
\caption{Lorentz factor of magnetized (black solid, $\sigma = 10^3$), and unmagnetized (red dotted)
outflows, as a function of radius.  Compactness of radiation source at $x=1$ is $\chi_s = 10^2-10^7$ (bottom to top
on the right side).}
\label{fig:gammabetacomparison}
\vskip .2in
\end{figure}

\section{Results}\label{s:results}

As the initial radiation compactness $\chi_s$ is increased, the plasma accelerates faster at small radius, 
and reaches a higher terminal Lorentz factor.  The radiation field is weakly collimated
near the emission radius, and so the starting Lorentz factor is reduced at 
large $\chi_s$.  Figure \ref{fig:gammabeta} shows the dependence of
$\Gamma$ and $\beta_{\phi}$ on radius, obtained for a strongly magnetized outflow
($\sigma=10^3$), starting from 10 times the light cylinder radius ($\omega = 10$).  
The Lorentz factor profile shows a smooth transition from Michel's minimum-energy 
solution in the low-$\chi_s$ limit, to the purely radiation-driven solution at large $\chi_s$.
An explicit comparison of magnetized and unmagnetized flows is made in Figure \ref{fig:gammabetacomparison}.
Once sees magnetization at a level $\sigma = 10^3$ makes little difference to $\Gamma(x)$ for $\chi_s \ga 10^5$.

The reduction in the starting Lorentz factor by radiation drag is tied to the negative inertia of the plasma inside
the fast critical point (Section \ref{s:fcrit}).  In this inner zone, the inertia is dominated by the magnetic field.  A
negative radial force, as is provided by the radiation field when $\Gamma>\Gamma_{\rm eq}$, pushes the flow to higher speeds.  
The situation reverses at the fast point, where matter begins to dominate the inertia and $\Gamma$
drops below $\Gamma_{\rm eq}$.  Outside the fast point, radiation provides a positive push on the matter
and the entrained magnetic field.  The asymptotic Lorentz factor grows
with respect to the cold MHD flow, as we discuss in more detail in Section \ref{sec:asympgamma}.  

\begin{figure}[h]
\centerline{\includegraphics[width=0.62\hsize]{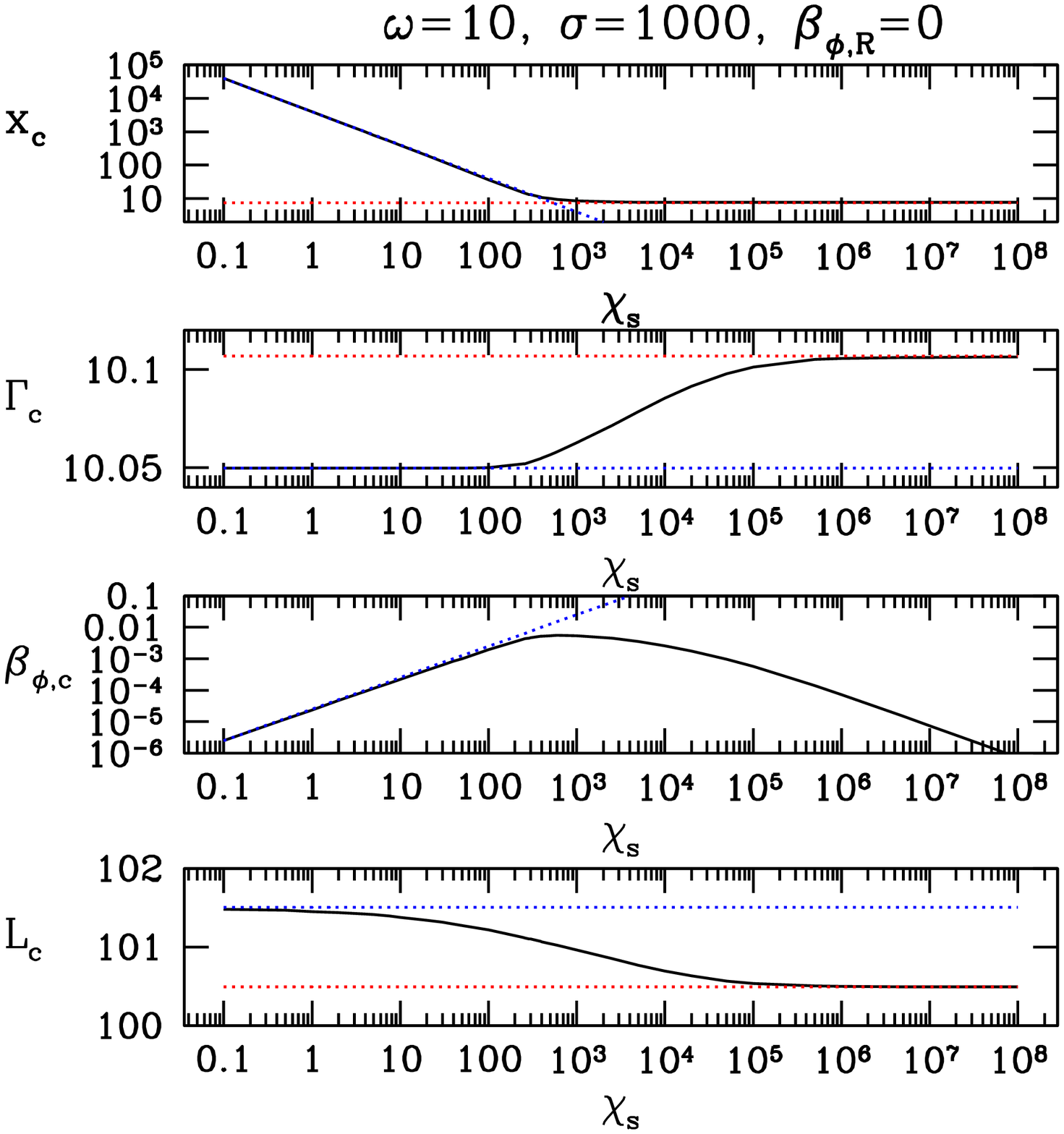}}
\vskip .2in
\centerline{\includegraphics[width=0.62\hsize]{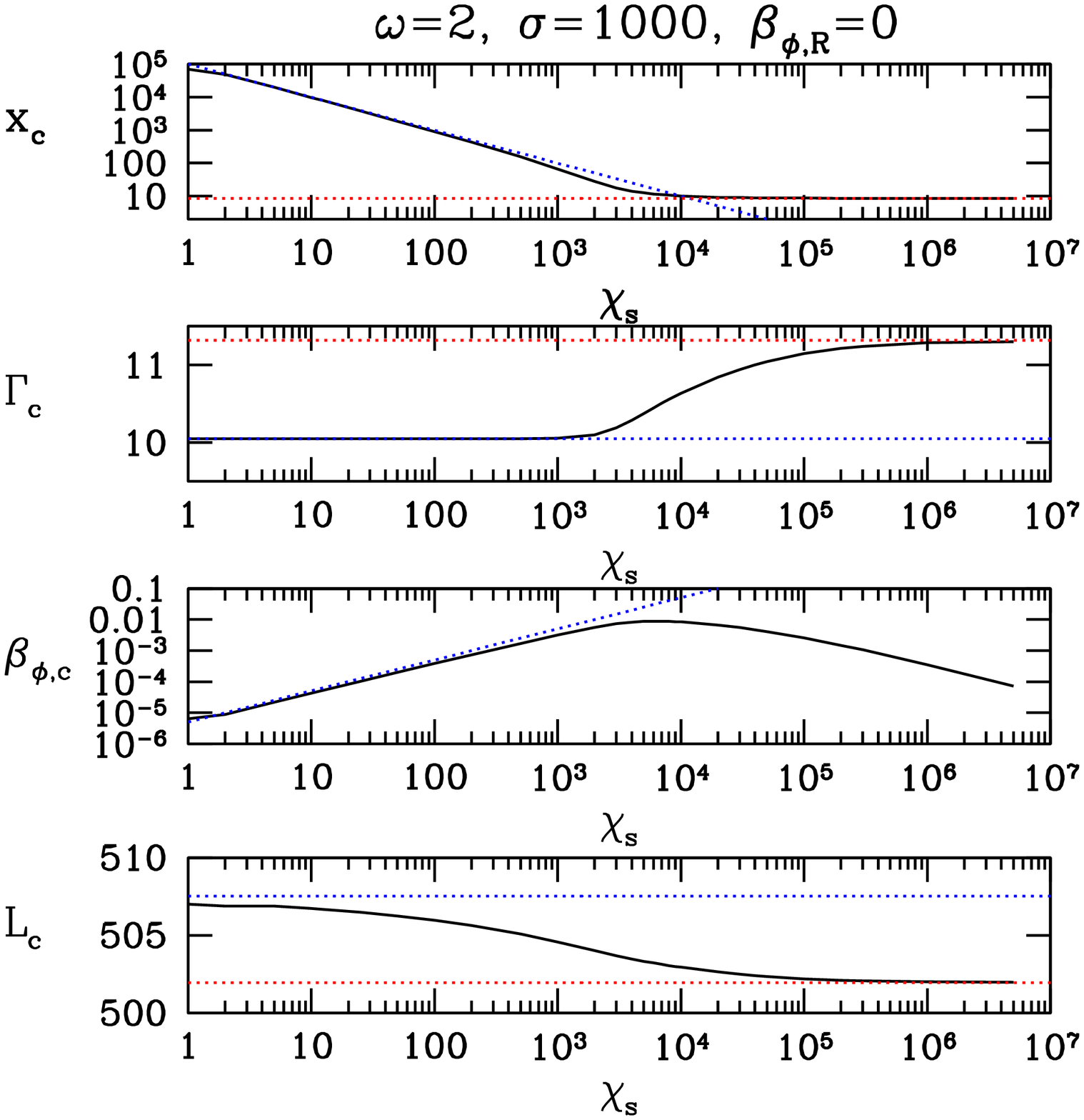}}
\caption{\emph{Top panel:} Radius $x_c$ of the fast critical point, and corresponding flow variables $\Gamma$, $\beta_{\phi}$ and 
$\mathcal{L}$, as a function of radiation compactness $\chi_s$, for $\sigma=10^3$, $\omega=10$.  Red and blue 
dotted lines show the analytical approximations (\ref{eq:lowcomp}) (low $\chi_s$) and (\ref{eq:highcomp}) (high $\chi_s$).
\emph{Bottom panel:}  Now with a smaller launching radius, $\omega=2$.}
\label{fig:criticalpoints}
\end{figure}

The fast critical point (marked by the red cross in Figure \ref{fig:gammabeta}) 
sits at infinity in the cold MHD wind with monopolar magnetic field, but moves in rapidly as 
$\chi_s$ increases above 10-100.  
The explicit dependence of the flow properties at the critical point on $\chi_s$ is shown in 
Figure \ref{fig:criticalpoints} for $\sigma = 10^3$, and two launching radii ($\omega = 10,2$).
Expression (\ref{eq:lowcomp}) provides an excellent fit to the
fast critical radius $x_c$ for $\chi_s \la 10^3$, above which it settles to the value where the equilibrium
Lorentz factor $\Gamma_{\rm eq} = \sigma^{1/3}$; see equation (\ref{eq:gameq}).  The radial field inertia
causes a small upward adjustment in $\Gamma_c$, as may be seen by comparing the cold MHD expression
$\Gamma_c \simeq \sigma^{1/3}$ with equation (\ref{eq:highcomp}).  There is a small downward adjustment
in the total angular momentum at the critical point.  

The coupling of the radiation field to the magnetic field has some subtle effects on the rotation of the 
outflow.  The angular velocity at the critical point
decreases with increasing $\chi_s$ at large compactness, as would be expected from the increasing
friction imparted by radiation field.  (The radiation field is assumed not
to rotate;  the effect of rotation is considered in Appendix \ref{s:rotation}.)  However, 
$\beta_{\phi,c}$ increases with
$\chi_s$ at small compactness, due to the shrinkage in the critical point radius.   The finite
value of $\beta_{\phi,c}$ is also worth commenting on.
The outward acceleration of the flow requires a finite angular speed for the matter.  Strong radiation drag,
acting alone, would rapidly damp the matter rotation.  However, the radiation field enters indirectly
into equation (\ref{eq:betaprime}) for $d\beta_{\phi}/dx<0$ through the term proportional to $\sigma\chi_s$.
This cross term represents the reaction of the radiation on the Lorentz force.
It is generally positive and almost exactly cancels 
the azimuthal radiation drag. (We enforce this condition at $x=1$,
but the condition is nearly satisfied automatically at all radii.)  This near cancellation allows the matter to 
maintain a high enough azimuthal speed to reach the fast critical point.
A further interesting effect, evident in Figure \ref{fig:gammabeta}, involves the near constancy of $\beta_{\phi}$ at small
radius in radiation-dominated outflows ($\chi_s\gg\sigma$).  This is due to a near cancellation of {\it all} the terms in $d\beta_{\phi}/dx$.

\subsection{Exchange of Energy Between Radiation and Magnetofluid}
\begin{figure}[h]
\epsscale{1.05}
\plottwo{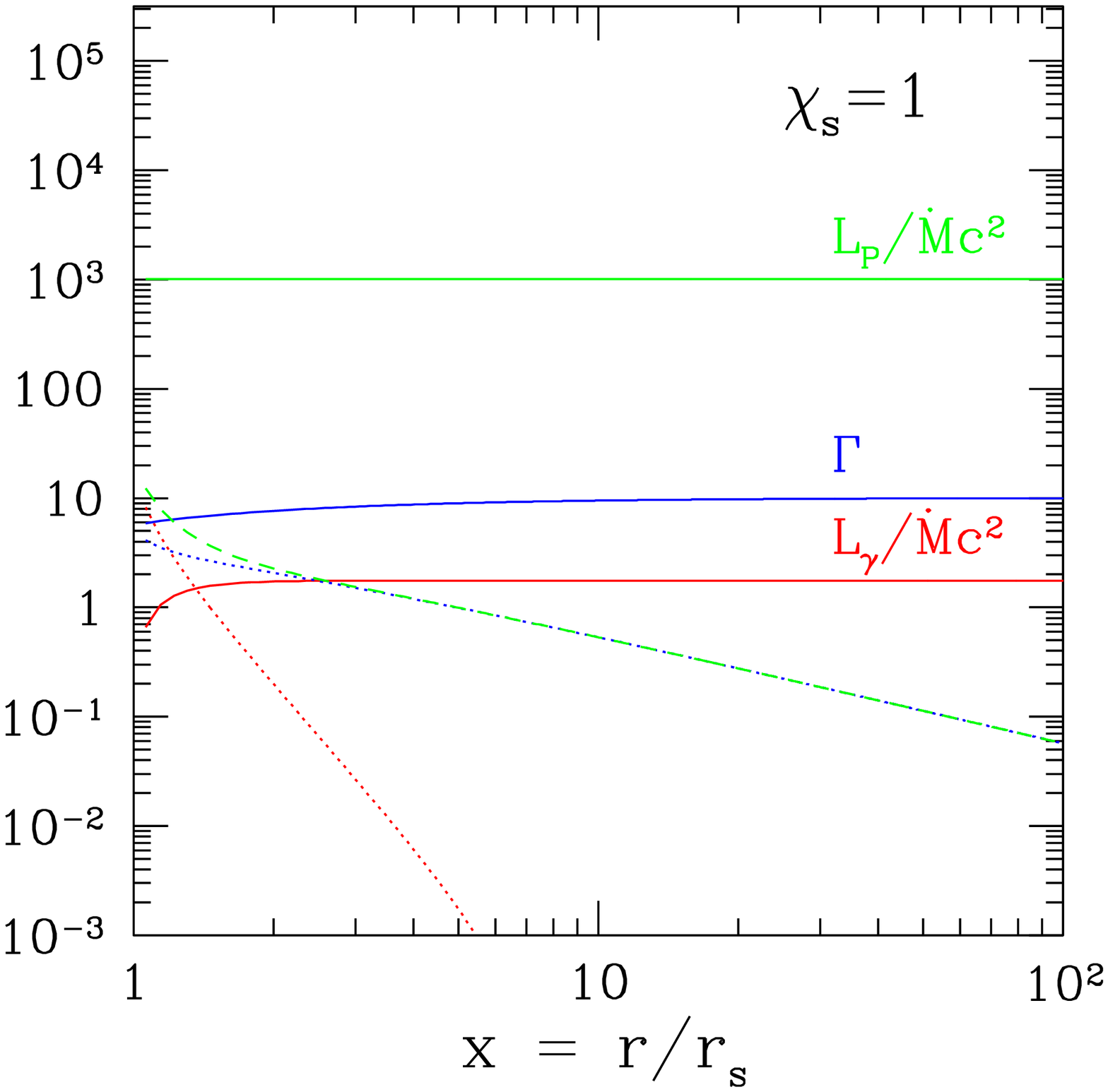}{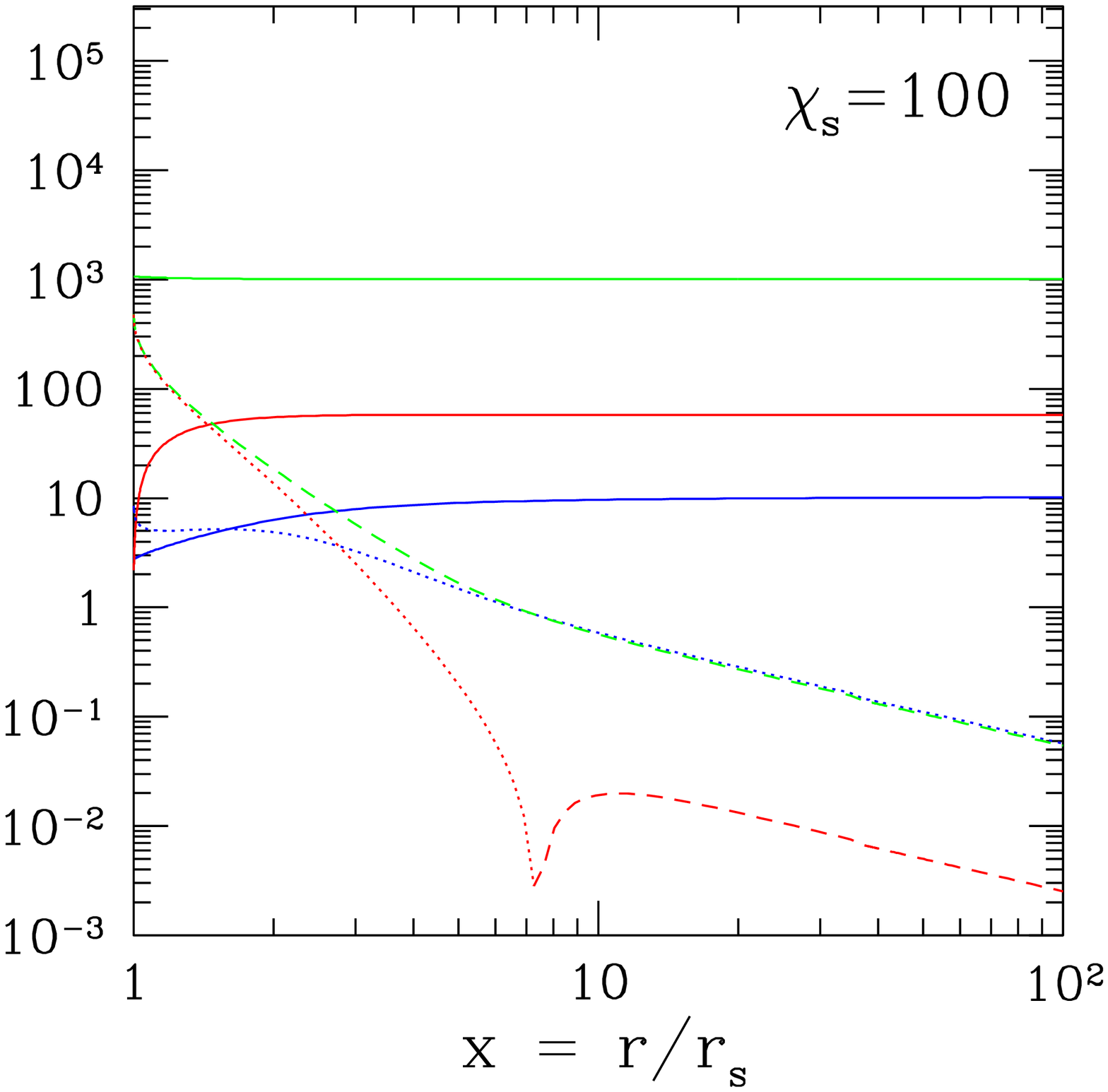}
\plottwo{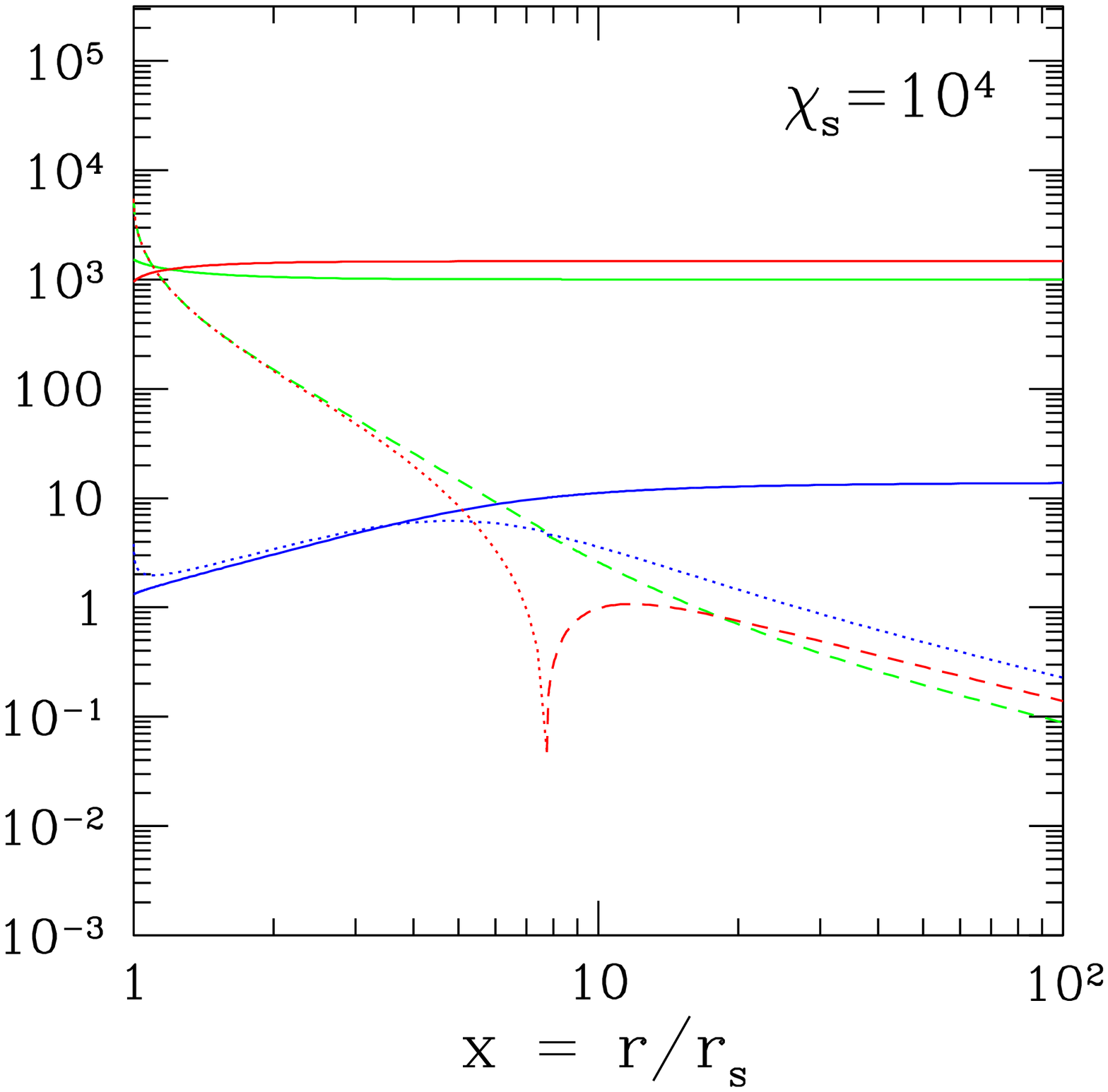}{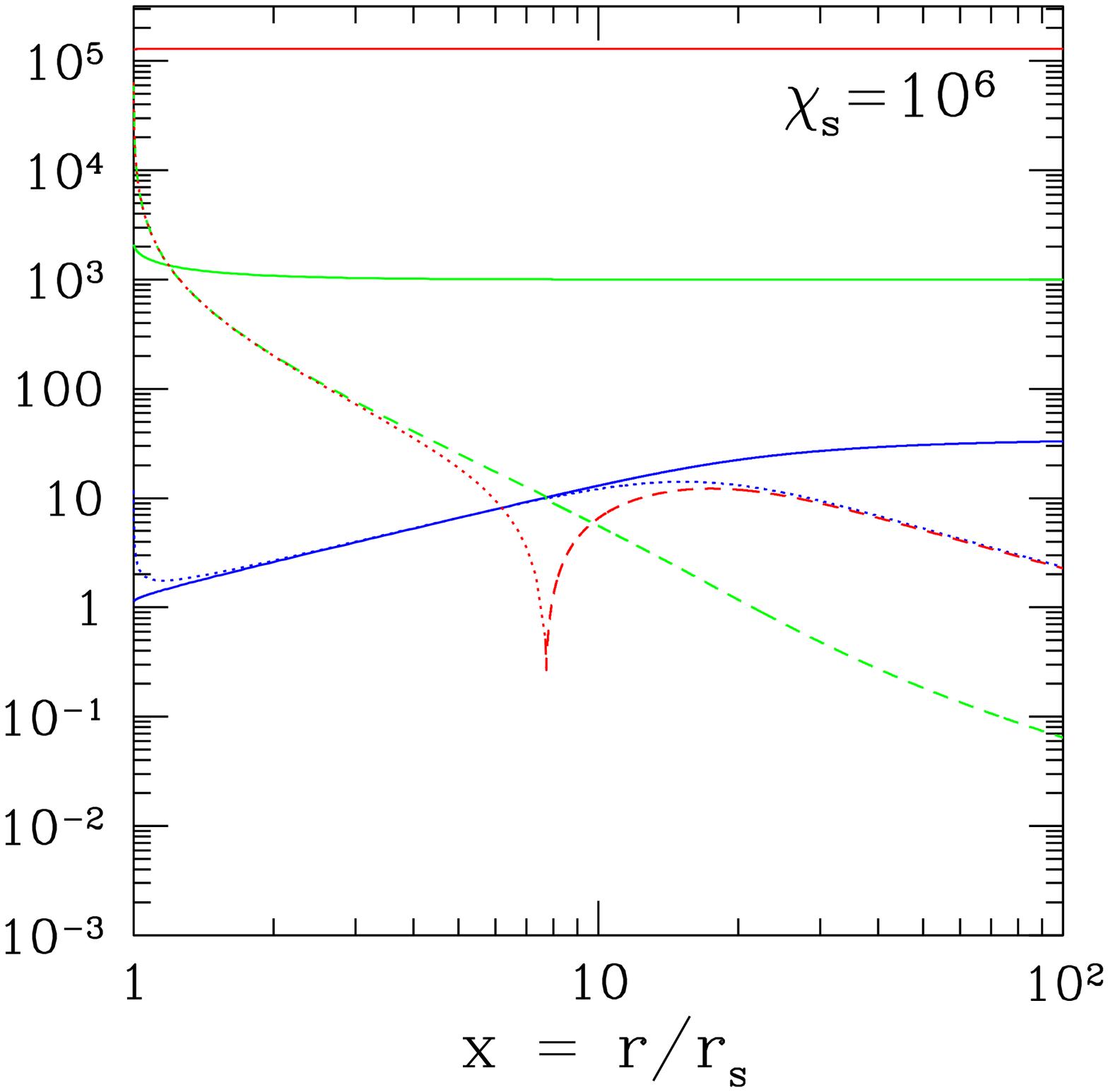}
\caption{Energy exchange in an outflow with $\sigma=1000$, $\omega=10$ with an optical depth of unity at $r=r_s$. Poynting, kinetic and radiation luminosities are plotted as solid lines and $dL/d\ln r$ for each is shown in corresponding colors with dotted/dashed representing positive/negative slopes.}
\label{fig:energyA}
\vskip .2in
\end{figure}

In a hot magnetized outflow, the fluid acts to couple the radiation and magnetic field allowing energy to be exchanged between them as the flow accelerates. To study this exchange we write the total energy in terms of kinetic, Poynting and radiation luminosities as
\be 
\frac{L_{K}+L_{P}+L_{\gamma}}{\dot{M}c^{2}}
= \Gamma - \frac{1}{x\omega\sin\theta}\frac{B_{\phi}}{B_{r}}\sigma + \frac{x\chi(x)}{6\tau_{{\rm es,s}}\Gamma_{s}^{2}}
\ee
where we have taken $\Gamma\propto r$ in calculating the optical depth. For simplicity, our solutions (being calculated in the optically thin
regime) assume a constant radiation flux and thus do not strictly conserve energy. Nevertheless, we can impose conservation using the obtained profiles for the kinetic and Poynting fluxes to study the qualitative features of the energy exchange. The results are shown in Figure \ref{fig:energyA} for various values of photon compactness while taking the optical depth at $r=r_s$ to be unity. The radiation luminosity generally increases sharply at small radius where photons are upscattered by the highly relativistic flow. This must be accompanied by a corresponding decrease in the magnetic energy since the Lorentz force is still accelerating the flow. To see why this occurs note that
well outside the light cylinder the Poynting luminosity in a monopolar outflow is $L_P\sim\beta_r^{-1}\sigma\dot{M}c^2$ which can decrease significantly only if the flow is not yet extremely relativistic at the photosphere. In this case field lines are coiled tight and transfer their magnetic energy to the radiation field as they unwind. The matter acts mainly as a catalyst in this process, with relatively small changes in kinetic energy. 

Outside the fast point the Poynting flux is essentially constant and further changes in the kinetic energy come directly from the interaction with the photon field. (In Figure \ref{fig:energyA} the positive/negative slopes of the energy profiles are plotted as dotted/dashed the corresponding colors.) When $\chi$ is high, the fast point marks a sharp transition in the energy exchange: conversion of magnetic to radiation energy inside; conversion of radiation to kinetic energy outside. Accounting for the increase in photon luminosity at small radius would enhance the acceleration outside the fast point.
 
\subsection{Application to Gamma-ray Burst Outflows with Displaced 
Photospheres}\label{s:dispphot}

We briefly consider the application of our results to gamma-ray bursts.  When 
a jet emerges from a Wolf-Rayet star, or from a cloud of neutron-rich debris, much of the
jet material may experience a strong outward Lorentz force \citep{tchek09}.  Even in that
circumstance, the radiation pressure force can enhance or impede the acceleration of the jet
material outside its photosphere (Russo \& Thompson 2012, Paper II).   Other parts of the jet
may have unfavorably curved flux surfaces and feel a weaker Lorentz force, more 
typical of the poloidal field geometry examined in this paper.

Although our solutions are found in the zone exterior to a static, photon-emitting surface, 
the photon field that emerges from a displaced photosphere will
have a similar effect on the outflow exterior to it, if we define an effective
source radius $r_{s,{\rm eff}}$ as in Section \ref{s:relphot}.  A high
photon intensity also has the effect of driving the flow profile to a linear relation
$\Gamma(r) \propto r$ inside the fast critical point.  In that case, the flow profile
does not depend on whether the outflow is optically thick or thin inside the critical
point.  The conditions for this to be the case are outlined at the end of
Section \ref{s:unmag}.

To fix some numbers, consider a gamma-ray outflow with an (isotropic) luminosity $4\pi dL_\gamma/d\Omega = 10^{51}\,L_{51}$ 
erg s$^{-1}$ and a photosphere of radius $r_\tau = 10^{10}r_{\tau,10}$ cm.  The corresponding photospheric compactness (\ref{eq:chitau}) 
is $\chi(r_\tau) = 1\times 10^8 L_{51}r_{\tau,10}^{-1}(m_p/\bar m)$.  Transparency at $r \lesssim r_\tau$ is guaranteed if 
$\Gamma(r_\tau) \gtrsim 10$, and pairs have mostly annihilated.  The effective source radius, obtained by equating the bulk Lorentz factor 
(\ref{eq:gameq}) of the radiation field with $\Gamma(r_\tau)$, is $r_{s,\rm eff} \sim r_\tau/\Gamma(r_\tau) < 10^9r_{\tau,10}$ cm.
The compactness scaled to this radius, which determines the amplitude of the radiation force in the wind equations 
(\ref{eqn:Gamp})-(\ref{eq:mueff}), is $\chi_s \gtrsim 10^9L_{51}r_{\tau,10}^{-1}[\Gamma(r_\tau)/10](m_p/\bar m)$.  

\subsection{Asymptotic Lorentz Factor:  Fully Transparent Outflows}\label{sec:asympgamma}

Since we are considering a strictly monopolar poloidal magnetic field, the asymptotic Lorentz factor is
limited to $\Gamma_{\infty,\sigma} \simeq \sigma^{1/3}$ at small radiation compactness.  
Higher Lorentz factors are possible at large $\chi_s$, where
acceleration is dominated by radiation pressure.  First consider the case where the outflow is fully transparent at $r \sim r_s$,
and moves transrelativistically near the inner boundary.  Then $\Gamma_{\infty,\chi} \simeq 1.086\chi_s^{1/4}$
(Section \ref{s:unmag}).  In the example at the end of the preceding Section, this corresponds to
$\Gamma_{\infty,\chi} \sim 200$.

\begin{figure}[h]
\epsscale{1.1}
\plottwo{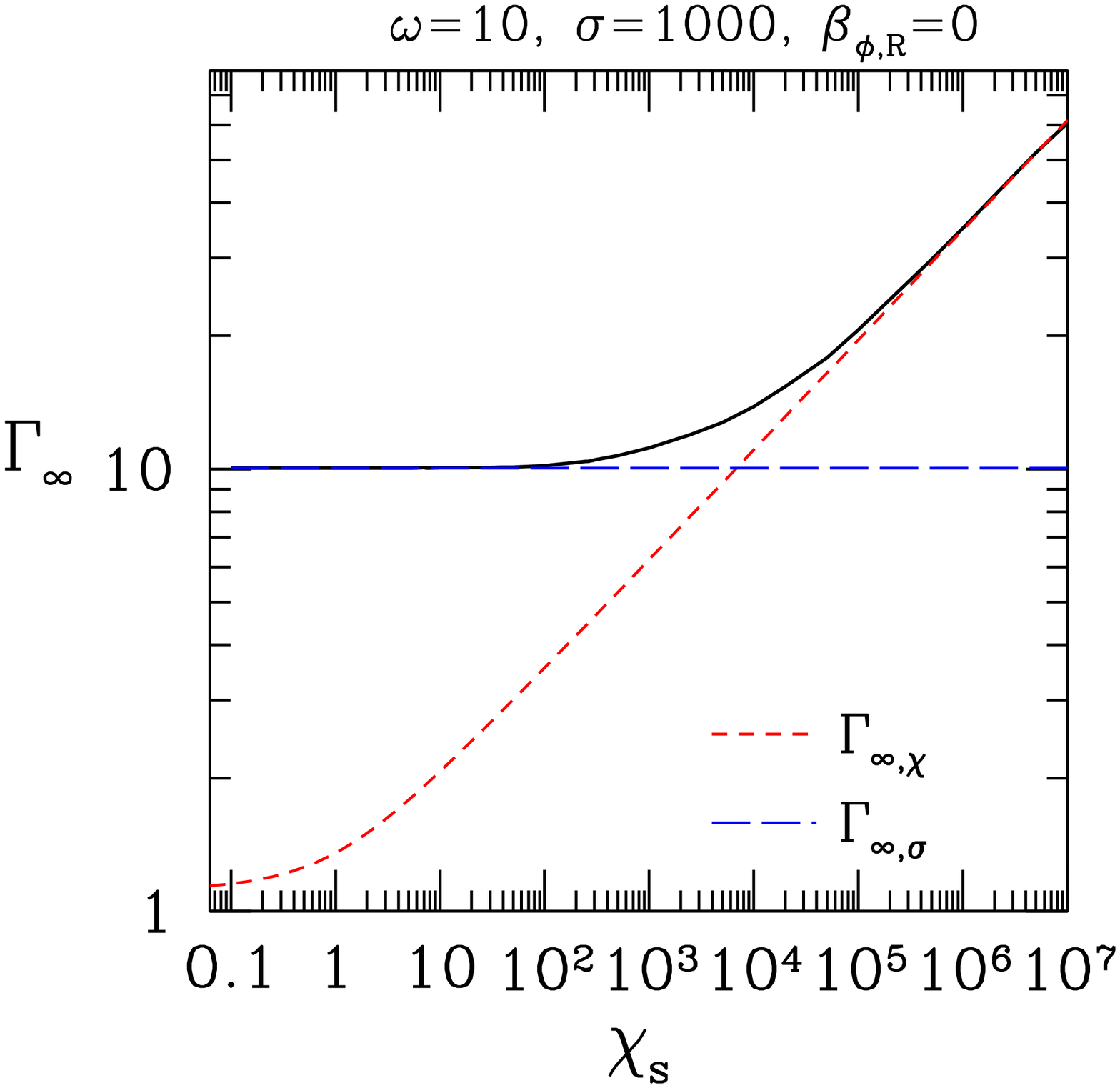}{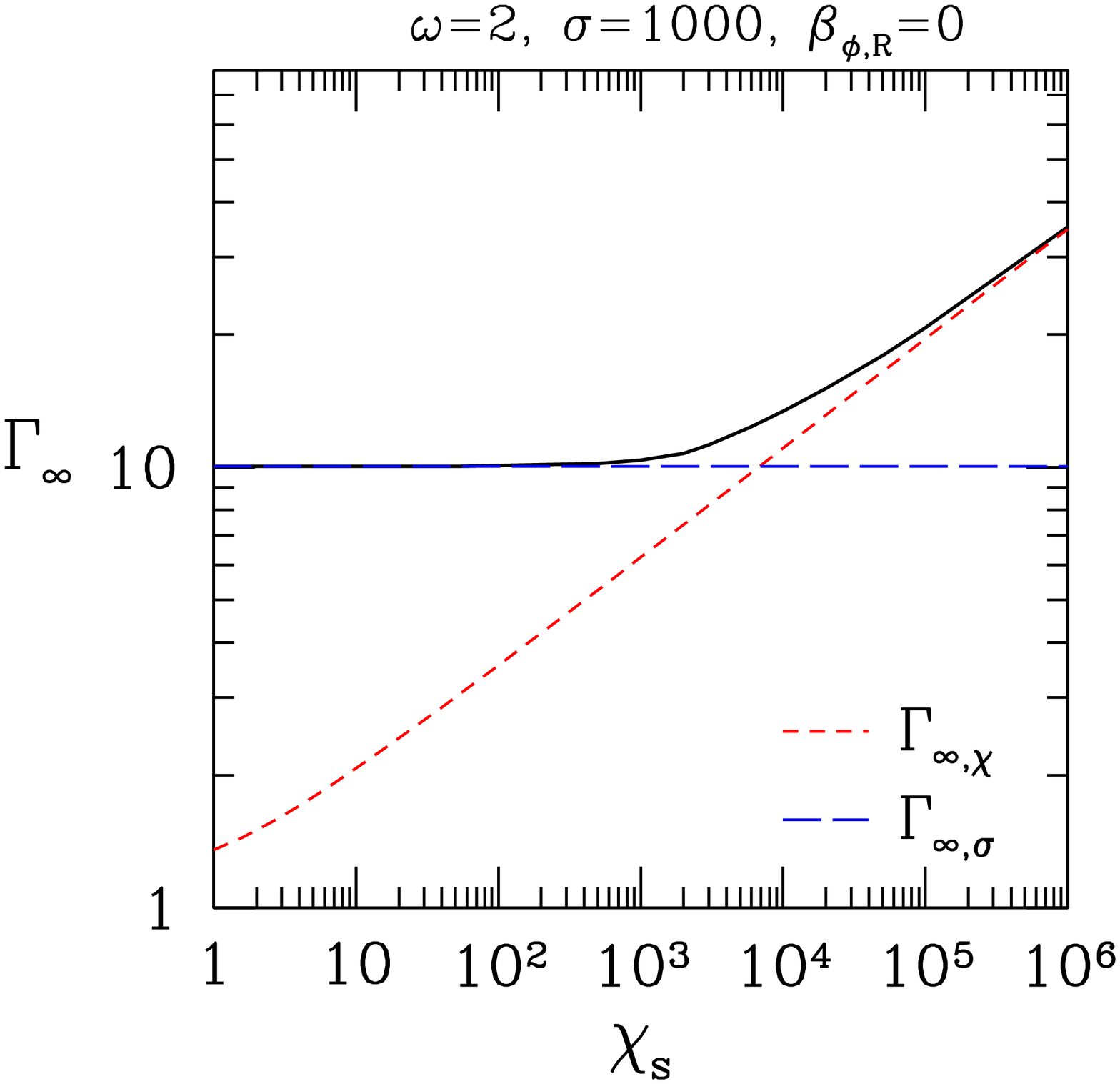}
\caption{Asymptotic Lorentz factor of a magnetized outflow ($\sigma=10^3$) exposed to a central radiation field
with compactness $\chi_s$ at $x=1$.  \emph{Left panel:} $\omega=10$. \emph{Right panel:} $\omega=2$}
\label{fig:gammainf}
\vskip .2in
\end{figure}

We find a smooth transition between these two limits as shown in Figure \ref{fig:gammainf} for 
$\sigma=10^3$ and $\omega=2,10$. This is well described by the function
\be\label{eq:gammainffit}
\Gamma_\infty^n =
\Gamma_{\infty,\sigma}^n + \Gamma_{\infty,\chi}^n,
\ee 
with $n\simeq 2.5$.  In this situation, strong radiative acceleration requires a photon source
that is not sourced internally by the outflow.  Otherwise, the photon luminosity is approximately
bounded above by the Poynting luminosity, which implies  $L_\gamma \lesssim L_{\rm P}$ and $\chi_s \la \sigma$.

\section{Spectrum of Scattered Photons}\label{s:spectrum}

The outflowing matter scatters the radiation field, and the frequency distribution
of scattered photons is non-thermal.   The effect grows stronger as the seed photon
beam grows wider with respect to the Lorentz cone, $\theta_s > 1/\Gamma$.  We now
calculate the spectrum in a situation where most photons see a low optical depth
after scattering by the magnetofluid.  

To motivate our calculation, it is worth discussing how such a situation could
arise in the context of GRBs.   Photons which start with a nearly blackbody spectrum
will maintain that spectral distribution in a simple relativistic fireball (e.g. \citealt{beloborodov11,lazzati11}).  
At the photosphere, the photon beam has an opening angle $\theta_s \sim \Gamma^{-1}$ 
and maintains a nearly isotropic distribution in the bulk frame.
So we expect minimal frequency redistribution in MHD outflows that are accelerated
mainly by radiation pressure, $\Gamma \sim \Gamma_{\rm eq}$.

The outflow feels a very strong Lorentz force where neighboring magnetic flux surfaces
diverge from each other, and we show in Paper II that this can driven $\Gamma > \Gamma_{\rm eq}$
even close to the photosphere of a relativistic outflow.  But even if this effect is
not operating, there is still good reason to expect that material from the progenitor star 
(in the case of collapsars), or from a preceding neutron-rich wind (in the case of
binary NS mergers) will interact with a relativistic MHD outflow.
This material can be massive enough to maintain a Lorentz factor $\Gamma_{\rm slow}$ much
lower than that of the MHD outflow, and so broaden the photon beam into a cone of width 
$\theta_s \sim 1/\Gamma_{\rm slow}$.   For example, a precursor shell, trapped at the head
of a relativistic jet, becomes Rayleigh-Taylor unstable as it is pushed outward,  opening out zones of an angular size
$\la 1/\Gamma$ in which the jet material can flow freely \citep{thompson06}.  Another possibility is
a jet boundary layer containing material of an intermediate Lorentz factor.

It is, therefore, posssible to probe the spectral signature of slow material
in a GRB, independent of the details of photon creation at large optical depth.  Indeed,
because photons moving off the axis of a relativistic flow scatter at a higher rate, this
type of interaction can occur even {\it outside} the photosphere of the MHD outflow, 
and still result in significant rescattering by the faster material.   There are 
interesting implications for both the high- and low-frequency portions of the photon spectrum.

\subsection{Scattering of a Monochromatic Source with Uniform Intensity}

We are interested in the power radiated from a steady flow, with a fixed particle density
at a given radius.  In the case of an impulsive event such as a GRB, of duration
$\Delta t$, the flow is effectively steady near the transparency surface if
$\Gamma^2(r_\tau) \gg r_\tau/c\Delta t$.  In this situation, there is no correction
between the time coordinate of an observer sitting at a large (but non-cosmological) distance 
from the engine, and the time coordinate $t$ of the engine rest-frame.  

We start with a monochromatic photon source
$I_\nu = I_0\nu_0\delta(\nu-\nu_0)$, and the top-hat angular distribution (\ref{eq:thetas})
corresponding to a (virtual) emission radius $r_s$.   More general source spectra are then
considered by a convolution.  The condition of elastic (Thomson) scattering is
\be\label{eq:doppler}
\tildenuem \equiv {\nu_{\rm em}\over\nu_0}  \;=\;  {1-\beta\mu\over 1-\beta\mu_{\rm em}}
\;=\; {1+\beta\mu_{\rm em}'\over 1+\beta\mu'}.
\ee
The direction cosines of the incident and emitted photons are measured with respect to 
the radial direction (we neglect any small non-radial motion of the matter),
$\mu = \hat k\cdot\hat r$, $\mu_{\rm em} = \hat k_{\rm em}\cdot \hat r$.  They take the range 
\be\label{eq:murange}
\sqrt{1-{1\over x^2}} \leq \mu \leq 1; \quad\quad -1 \leq \mu_{\rm em} \leq 1.
\ee
Primed quantities are measured in the rest frame of a scattering charge.  

\begin{figure}[h]
\centerline{\includegraphics[width=0.52\hsize]{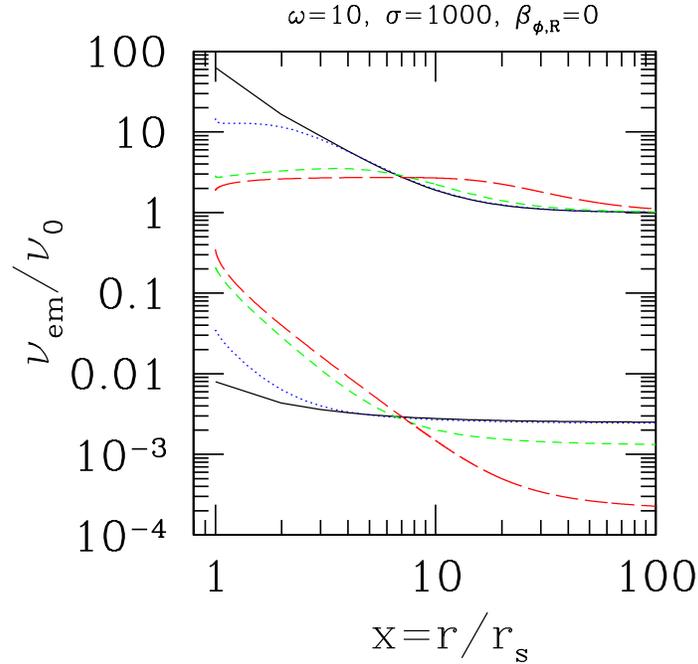}}
\caption{Maximum frequency of scattered photons (upper curves) and minimum frequency (lower curves),
as a function of radius, in a strongly magnetized spherical wind ($\sigma = 10^3$).
Compactness of photon source at $x=1$:  $\chi_s=1$ (solid black); $\chi_s=10^2$ (dotted blue);
$\chi_s=10^4$ (short-dashed green); $\chi_s=10^6$ (long-dashed red).}
\label{fig:nurange}
\vskip .05in
\end{figure}
\begin{figure}[h]
\centerline{\includegraphics[width=0.52\hsize]{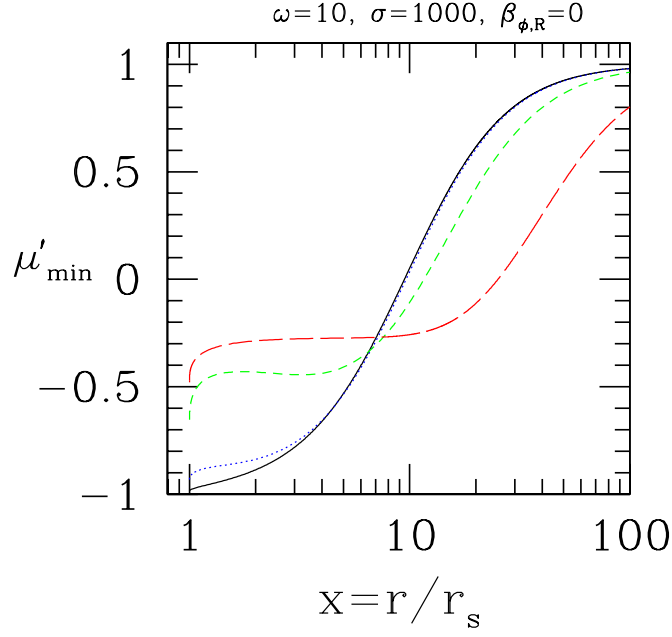}}
\caption{Asymmetry of the target photon distribution in the bulk frame of the outflow,
represented by the minimum direction cosine of the photons $\mu^\prime$, as
a function of radius.  $\mu^\prime$ is measured with respect to the flow direction, and
is larger than the plotted value at very low frequencies [equation (\ref{eq:mubounds})].
Compactness of photon source at $x=1$:  $\chi_s=1$ (solid black); $\chi_s=10^2$ (dotted blue);
$\chi_s=10^4$ (short-dashed green); $\chi_s=10^6$ (long-dashed red).}
\label{fig:murange}
\vskip .1in  
\end{figure}

A scattered photon reaches the maximum frequency
\be\label{eq:numax}
\nu_{\rm em,max} = {1-\beta\mu_{\rm min}\over 1-\beta}\nu_0 \simeq \left(1+{\Gamma^2\over x^2}\right)\nu_0
\quad (x,\Gamma\gg 1).
\ee
Hence the scattered spectrum  develops a significant tail at frequencies above $\nu_0$ if
the outflow has a Lorentz factor $\Gamma \ga \Gamma_{\rm eq}$ [equation (\ref{eq:gameq})].
This non-thermal tail is more pronounced at lower values of the compactness
$\chi_s$, where the Lorentz force dominates the acceleration of the flow.  The scattered
spectrum also extends to a low frequency $\nu_{\rm em,min} = [(1-\beta)/(1+\beta)]\nu_0 \sim
\nu_0/4\Gamma^2$.  Sample profiles of $\nu_{\rm em,max}$ and $\nu_{\rm em,min}$ are given in 
Figure \ref{fig:nurange}, and the asymmetry of the rest frame photon flux
in Figure \ref{fig:murange}.

The frequency $\nu_{\rm em}$ of the emitted photon is hardest in the
flow direction.  Softer photons are emitted off this axis, and also see a larger
optical depth to scattering.  In a spherically symmetric outflow, a photon emitted at
radius $r$ sees a scattering depth
\be\label{eq:tau0}
\tau_{\rm es}(r,\mu_{\rm em}) = {\sigma_T\over c}{d\dot N\over d\Omega}\
\int_r^\infty\left[1-\beta(r_2)\mu(r_2)\right] {dr_2\over \beta(r_2) \mu(r_2) {r_2}^2}.
\ee
The direction cosine evolves from the emission radius $r$ to $r_2 > r$
according to
\be
1 - \mu(r_2)^2 =  \left({r\over r_2}\right)^2(1-\mu_{\rm em}^2)
\quad\quad(r_2 > r).
\ee

It is straightforward to calculate the emergent photon spectrum by a Monte Carlo
method.  The direction cosines of the input photons are drawn randomly from the 
uniform distribution (\ref{eq:murange}) at $x=1$.  The optical depth $\Delta \tau_{\rm es}$
to the first (next) scattering is determined by randomly picking $1-e^{-\Delta\tau_{\rm es}}$, 
followed by a step-by-step integration of $\tau_{\rm es}$ along the ray.  Scattering is
performed in the rest frame of the cold flow, by transforming $\mu' = 
(\mu-\beta)/(1-\beta\mu)$, picking rest frame scattering angles $\theta_s'$,
$\phi_s'$ with respect to this axis, and then determining the direction cosine
of the outgoing photon via $\mu_{\rm em}' = \mu'\cos\theta_s' + (1-\mu'^2)^{1/2}\sin\theta_s'\cos\phi_s'$.
The photon escapes if $\Delta\tau_{\rm es}$ exceeds the total optical depth along the ray.
Working in spherical symmetry, we record the frequency but not the direction of the outgoing photon.


\begin{figure}[h]
\epsscale{1.17}
\plottwo{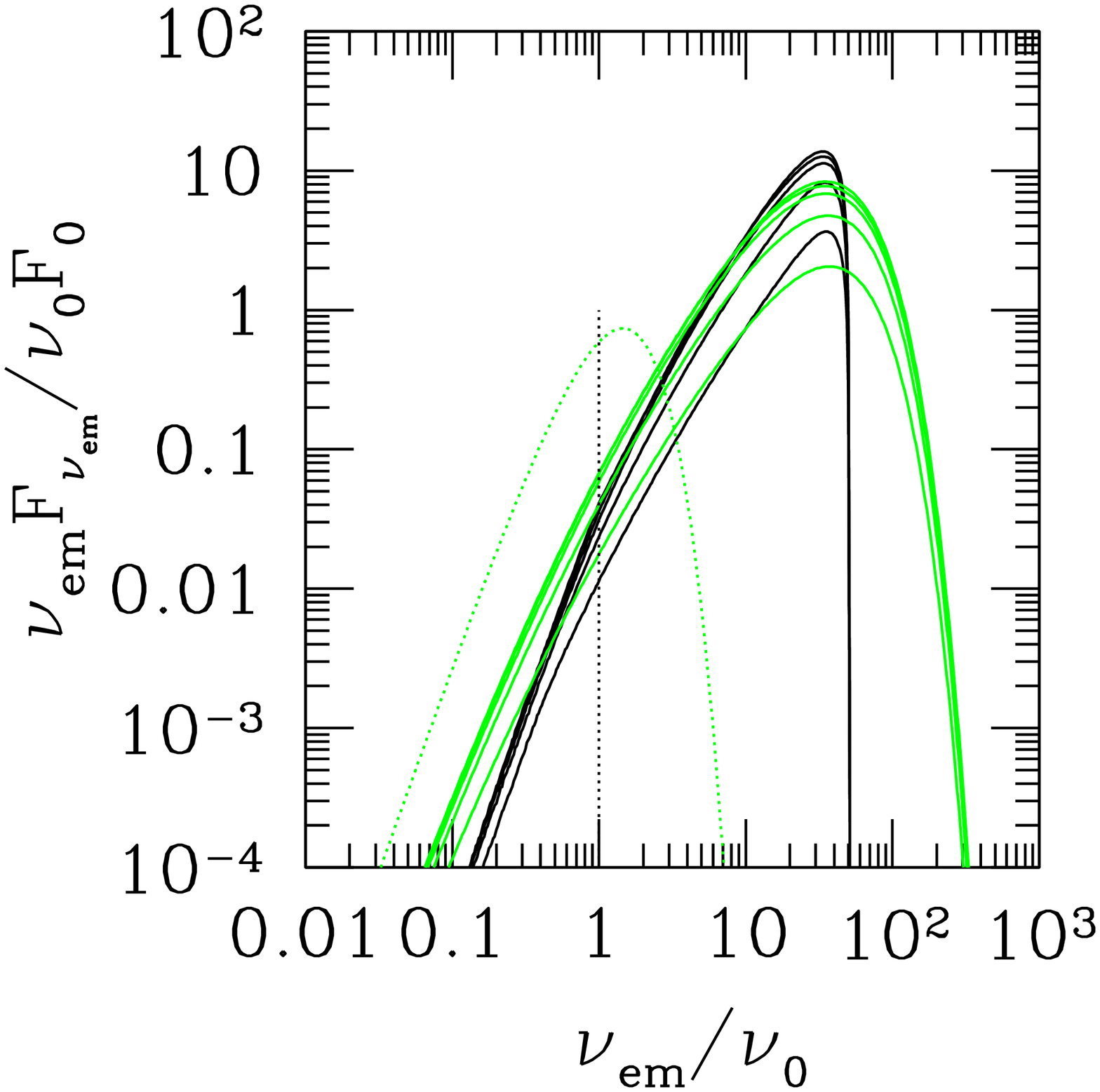}{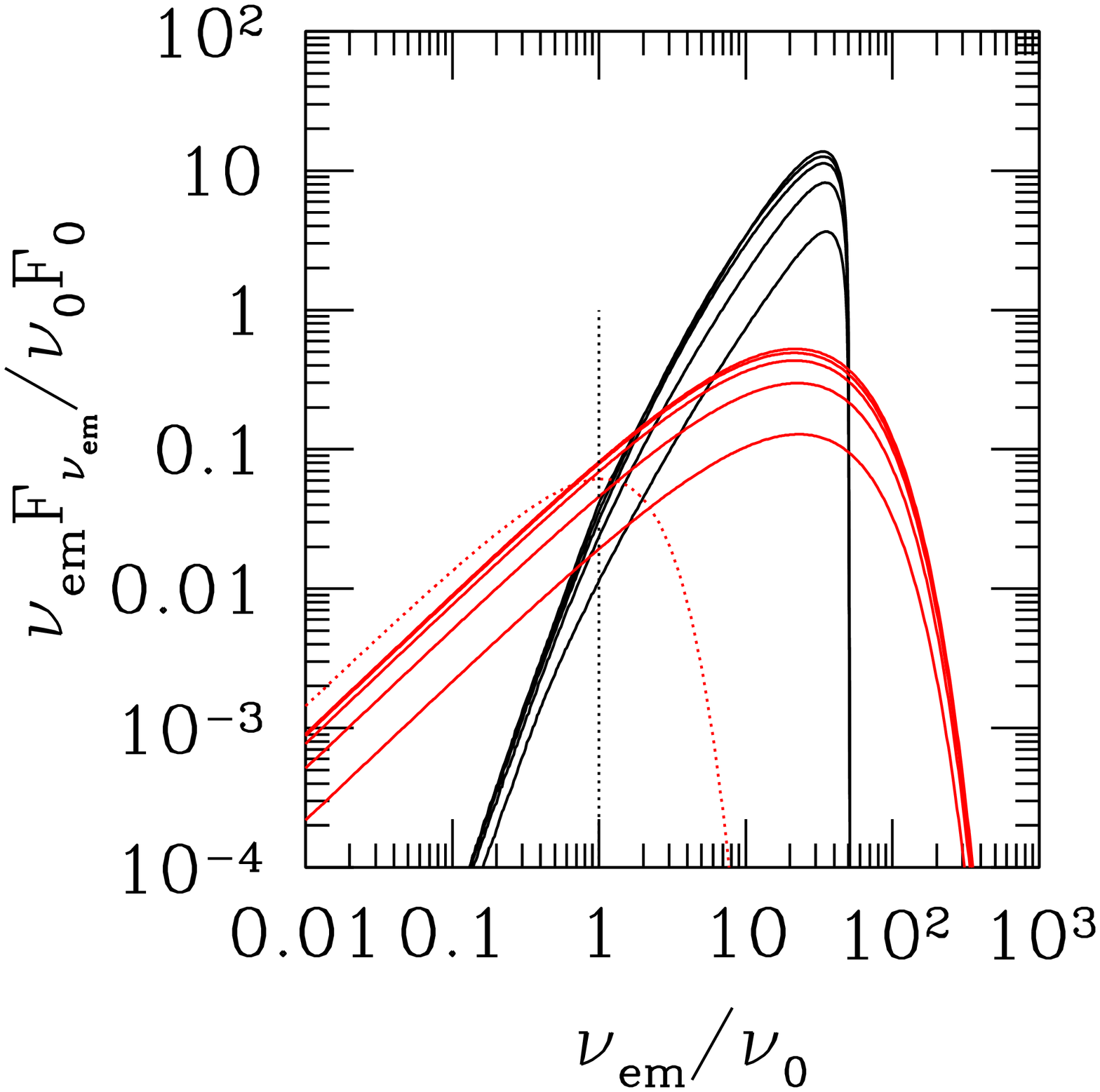}
\caption{Spectrum of scattered photons emerging from a spherical outflow with a simple profile $\Gamma(r) = 5\,(r/r_s)$, 
obtained by direct Monte Carlo integration.  Optical depth (\ref{eq:tauoffaxis}) of the off-axis photons is $0.3,1,3,10,30$
from bottom to top, corresponding to optical depths $\la 1$ after scattering at frequencies 
$\ga \nu_0$.  Unscattered source photons not included (see Figure \ref{fig:spectra0} for comparison).
Black lines:  monochromatic source spectrum. \emph{Left panel:}  
black-body photon source (green lines).  \emph{Right panel:}  GRB-like source spectrum (red lines),
$F_\nu \sim {\rm const} \times e^{-h\nu/kT_0}$ (the low-frequency half of the Band function extended to higher frequencies). 
In both panels, the source spectrum is the dotted curve.}
\label{fig:spectra00}
\vskip 0.85in
\end{figure}
\begin{figure}[h]
\epsscale{1.17}
\plottwo{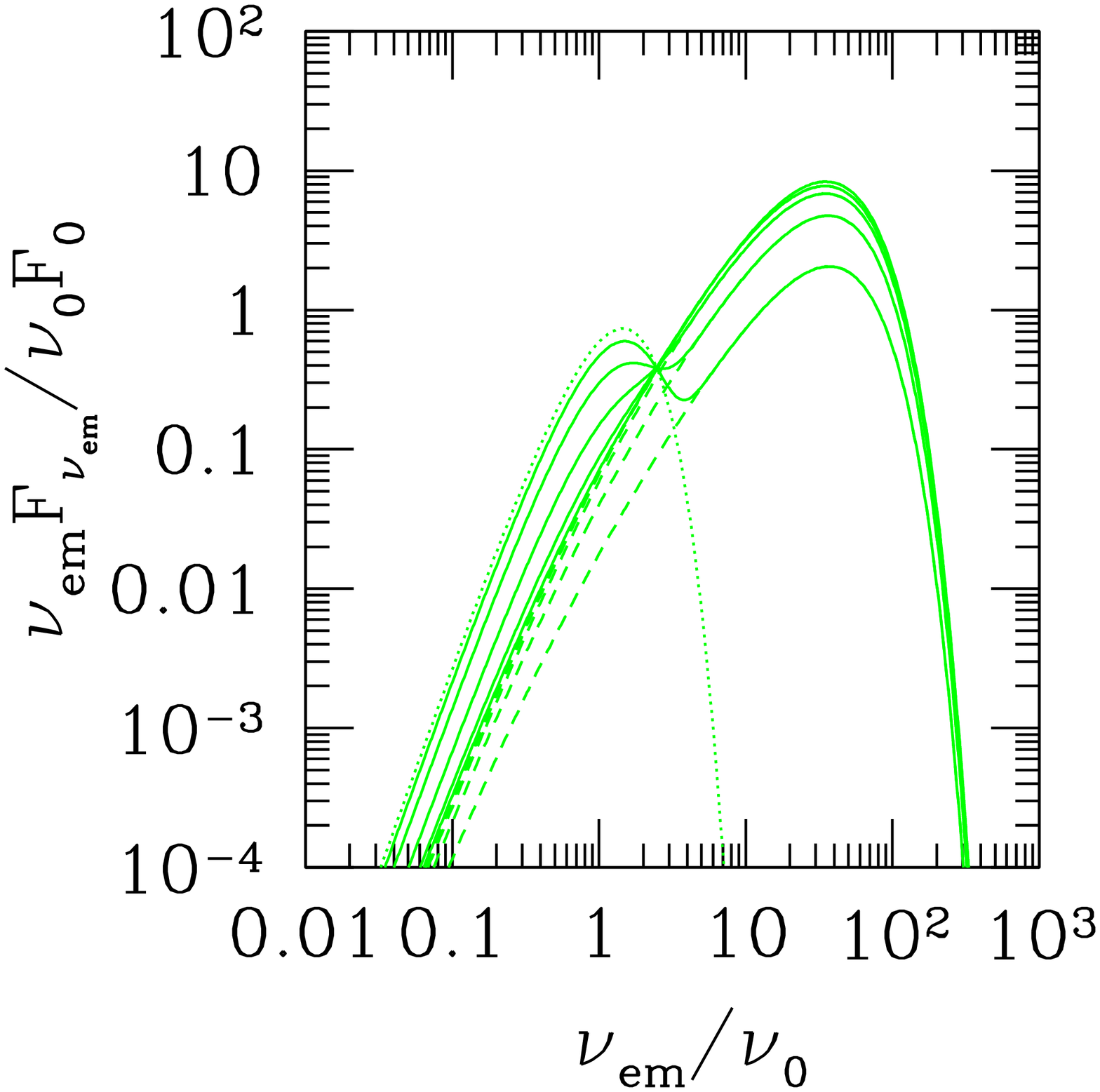}{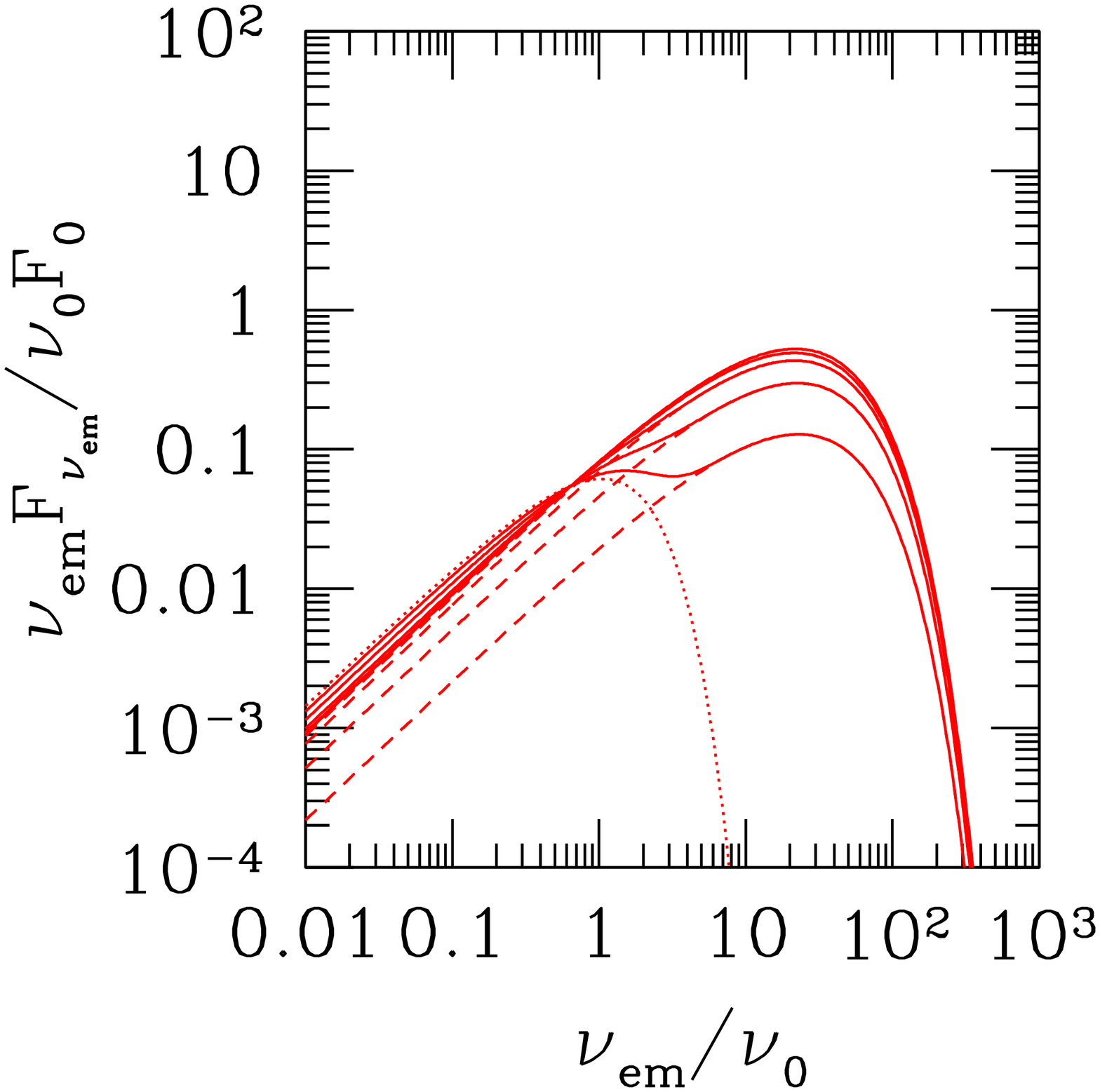}
\caption{Solid curves:  same as Figure \ref{fig:spectra00}, but adding in the unscattered component of the 
source spectrum.  The scattered spectrum dominates near the seed blackbody peak when the off-axis optical
depth $\tau_{\rm es}(r_s,\mu_{\rm min}) \ga 10$.  Dashed curves:  only scattered photons included.}
\label{fig:spectra0}
\vskip .2in
\end{figure}

\subsection{Linear Flow Profile}

To illustrate some of the main effects, we show in Figure \ref{fig:spectra00} the spectrum
resulting from an outflow with a simple linear profile $\Gamma(r) = 5\,(r/r_s)$,
and various values of the scattering depth experienced by the most obliquely propagating photons.
In this case, the bulk frame of the seed photons moves at a somewhat lower Lorentz factor than the
magnetofluid, $\Gamma_{\rm slow} \sim 0.2\, \Gamma$.   The scattered spectrum therefore peaks 
well above the seed frequency, $\nu_{\rm em,max} \sim 30\,\nu_0$ [equation (\ref{eq:numax})],
driven by bulk Comptonziation.

First consider the output from a monochromatic source spectrum (the black lines in Figure \ref{fig:spectra00}).
The output spectrum is fairly flat, $F_\nu \sim \nu^{0.5}$, over a decade in frequency below the peak, 
steepening to $F_\nu \sim \nu$ at $\nu \sim \nu_0$.  
The scattered photons see a small optical depth (suppressed by a factor $\sim (\Gamma_{\rm slow}/\Gamma)^2$
down to fairly low frequencies, where the slope approaches a Rayleigh-Jeans value.  

As the optical depth 
of the seed photons is increased, the spectrum steepens slightly at high frequencies.  
The spectra in Figure (\ref{fig:spectra00}) are labelled by the optical
depth seen by seed photons at the maximum angle $\theta_s \sim (r/r_s)^{-1}$ and minimum direction
cosine $\mu_{\rm min} \simeq 1-\theta_s^2/2$.  Requiring that $\tau_{\rm es}(r_s,\mu_{\rm min}) \ga 1$
allows the radial optical depth to remain small if $\Gamma\cdot\theta_s \gg 1$:
\be\label{eq:tauoffaxis}
\tau_{\rm es}(r_s,\mu_{\rm min}) \;\sim\; \left[1 + \Gamma^2(r_s)\right]\,\tau_{\rm es}(r_s,1)
\quad\Rightarrow\quad \tau_{\rm es}(r_s,1) \;\ga\; {1\over 1 + \Gamma^2(r_s)}.
\ee

\begin{figure}[h]
\epsscale{1.17}
\plottwo{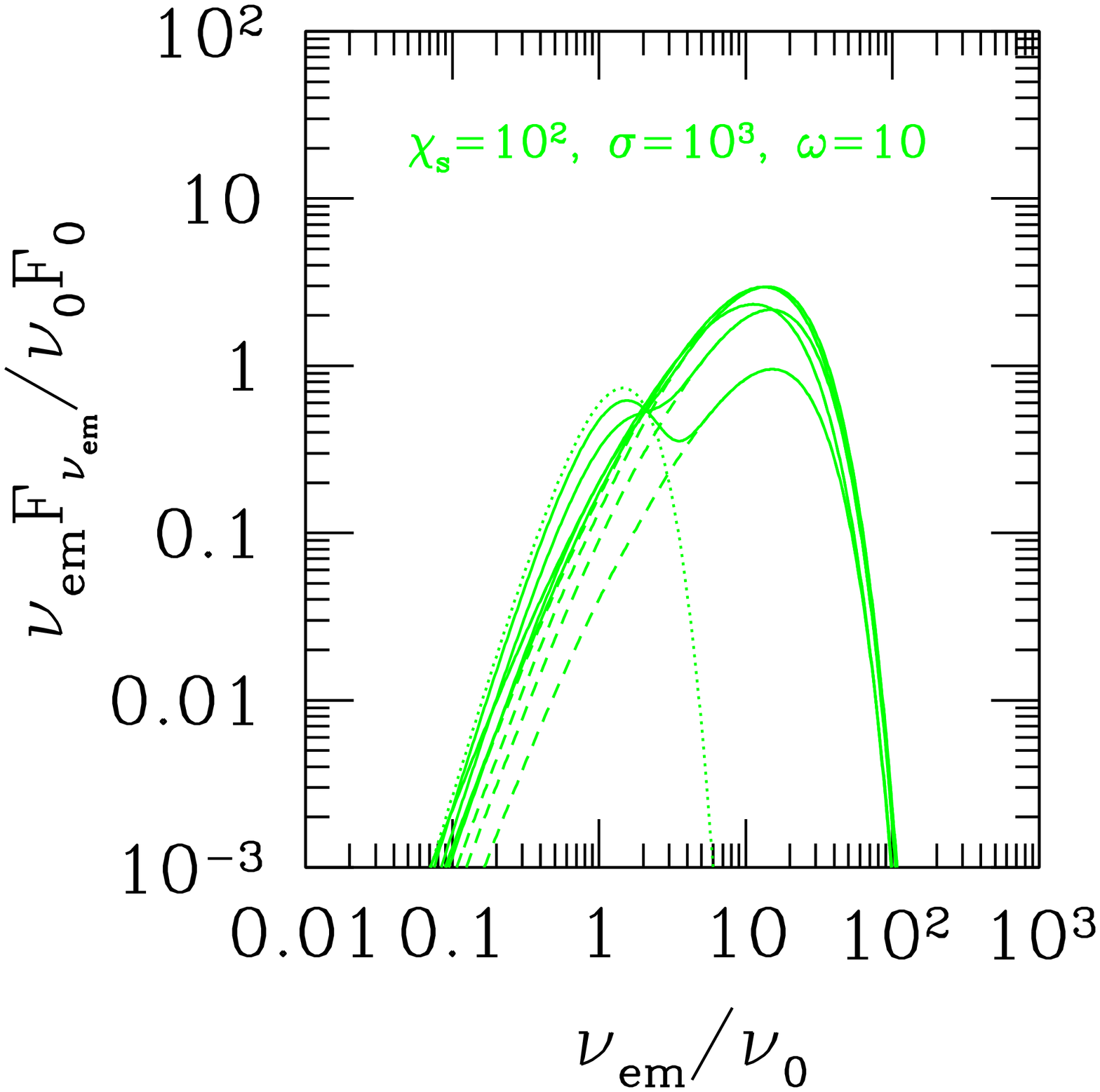}{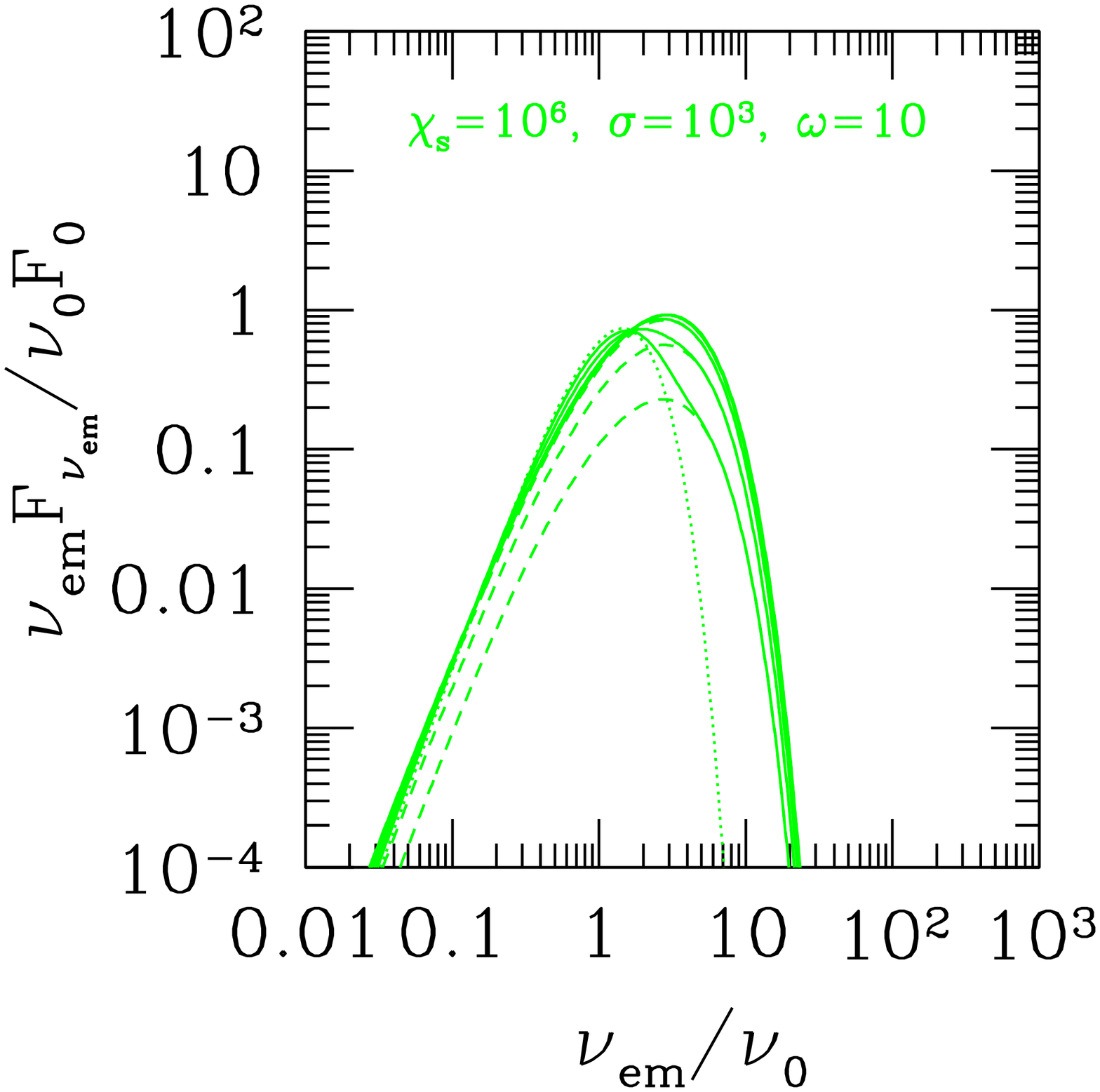}
\plottwo{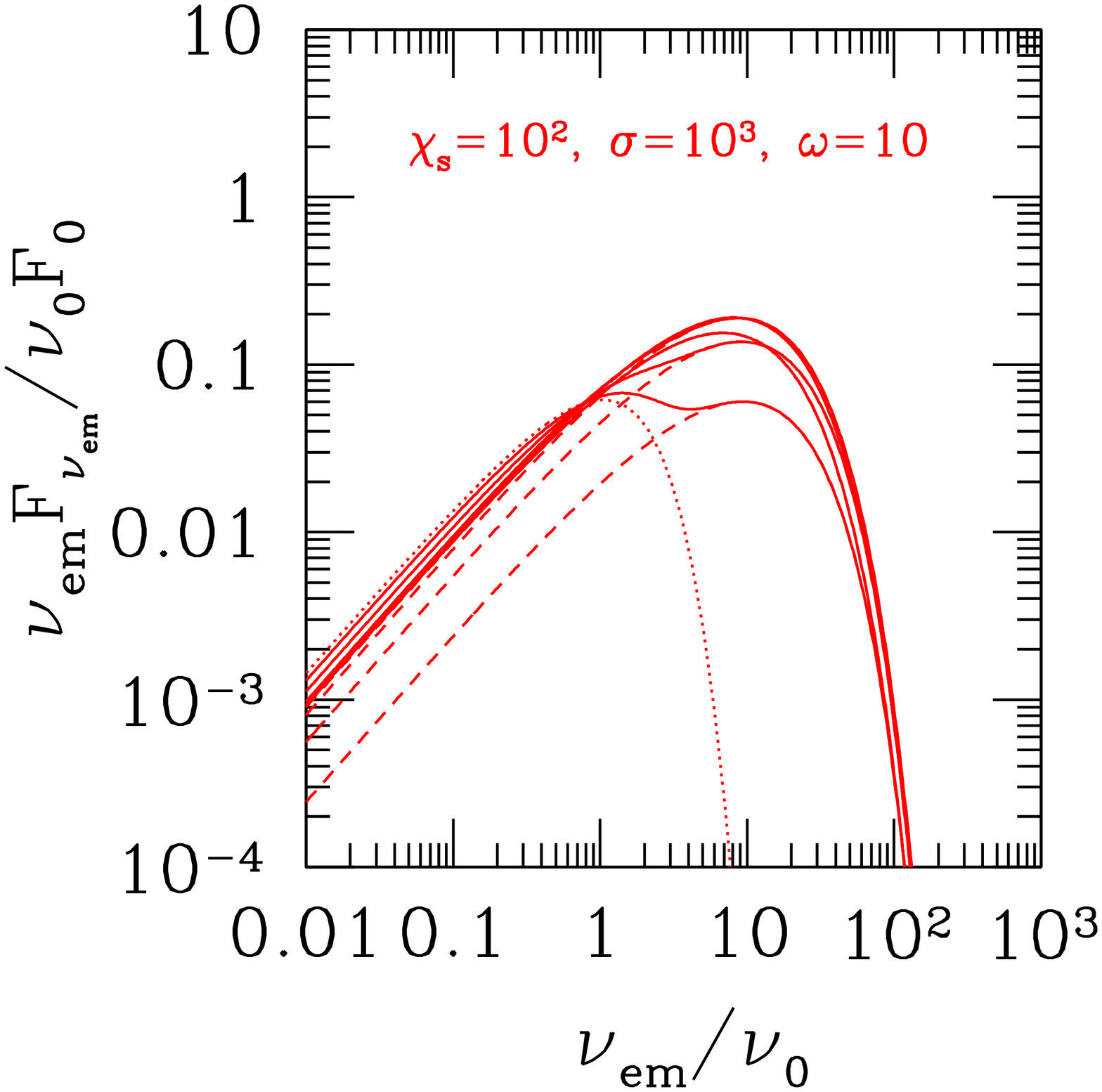}{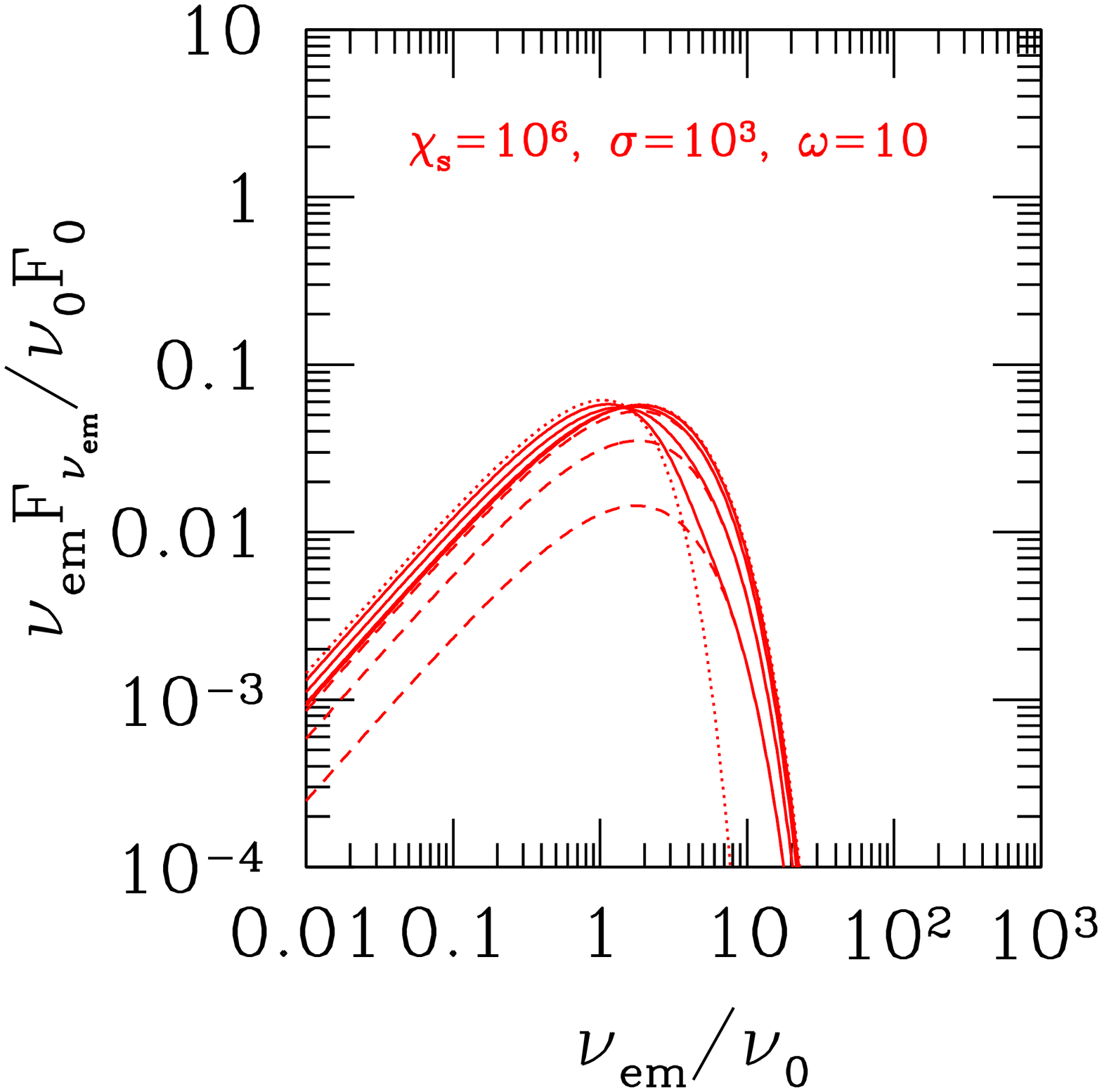}
\caption{Total spectrum of photons emerging from a hot electromagnetic outflow ($\sigma = 10^3$)
with a blackbody source of compactness $\chi_s = 10^2$ (left panel) and $\chi_s = 10^6$ (right panel).  
Flow profile shown in Figure \ref{fig:gammabeta}.  
Curves from bottom to top have off-axis optical depth (\ref{eq:tauoffaxis}) equal to $0.3,1,3,10,30$.
Increasing $\tau_{\rm es}(r_s,\mu_{\rm min})$ implies that the outflow moves closer to
the equilibrium value $\Gamma_{\rm eq}$ at the transparency radius, hence the steeper
high-frequency spectrum.  Dashed spectra do not include unscattered source radiation.}
\label{fig:spectra}
\vskip .1in
\end{figure}
\begin{figure}[h]
\centerline{\includegraphics[width=0.58\hsize]{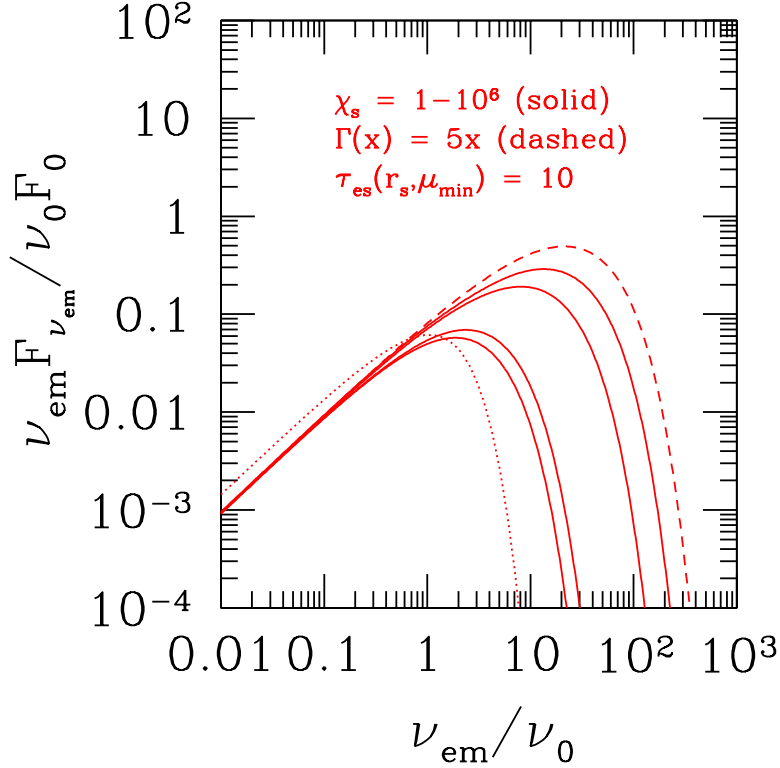}}
\caption{Same as Figure \ref{fig:spectra}, with a GRB-like
photon source (dotted curve), for a range of compactness $\chi_s = 1, 10^2, 10^4, 10^6$ (solid curves, softening
at high frequencies with increasing compactness).  Dashed line shows result for linear flow profile, $\Gamma(x) = 5x$.
Outflow has a fixed off-axis optical depth $\tau_{\rm es}(r_s,\mu_{\rm min}) = 10$. 
See Figure \ref{fig:gammabeta} for corresponding flow profiles.}
\label{fig:spectra2}
\end{figure}

\subsection{Generalization to Blackbody and Other Thermal Seeds}\label{s:bb}

It is straightforward to calculate the output spectrum for a broadband source with spectrum $F_{\nu 0}(\nu_0)$ via
\be
F_{\nu_{\rm em}}(\tildenuem) \rightarrow \int {F_{\nu_0}(\nu_0')\over F_0} F_{\nu_{\rm em}}\left({\nu_{\rm em}\over\nu_0'}\right) {d\nu_0'\over\nu_0'}.
\ee
Here $F_{\nu_{\rm em}}(\tildenuem)$ is the Greens function response (\ref{eq:fnusing}) of the scattering outflow
to a line photon source.

The result for a blackbody seed is shown in the left panel of Figure \ref{fig:spectra00}, and for
a thermal seed with a GRB-like spectrum at low frequencies, $F_\nu = {\rm constant}\times e^{-h\nu/kT_0}$, in the right panel.\footnote{This
corresponds to the low-frequency half of the Band function \citep{band93}.  The high-frequency power-law
tail of the observed GRB spectrum must then result from further upscattering by a process not considered in this paper.}  
In both cases, the seed temperature is normalized so that 
$F_\nu$ peaks at $\nu = \nu_0$.   In the second case, the upscattered portion of the spectrum at $\nu > \nu_0$ connects smoothly
with the seed spectrum at $\nu < \nu_0$. 

One can also add in the source spectrum, appropriately attenuated,
\be\label{eq:sourcetau}
F_{\nu 0} \;\rightarrow\; 2\pi\int I_{\nu 0} \exp\left[-\tau_{\rm es}(r_s,\mu)\right]d\mu 
\;\simeq\; {1+\Gamma^{-2}(r_s)\over\tau_{\rm es}(r_s,\mu_{\rm min})} 
\Bigl\{\exp\left[-\tau(r_s,1)\right]-\exp\left[-\tau(r_s,\mu_{\rm min})\right]\Bigr\}\,F_{\nu 0},
\ee
where we have made use of the small-angle approximation (\ref{eq:tau}) to the scattering depth.  The effect is
shown in Figure \ref{fig:spectra0}.  The seed blackbody peak is apparent if the seed photons see a 
maximum optical depth $\tau(r_s,\mu_{\rm min}) \sim 1$, and helps to extend the low-energy tail for 
$\tau(r_s,\mu_{\rm min}) \sim 3$.  It is subdominant for larger optical depths.

\subsection{Results for Radiatively Driven MHD Winds}

\begin{figure}[h]
\epsscale{1.15}
\plottwo{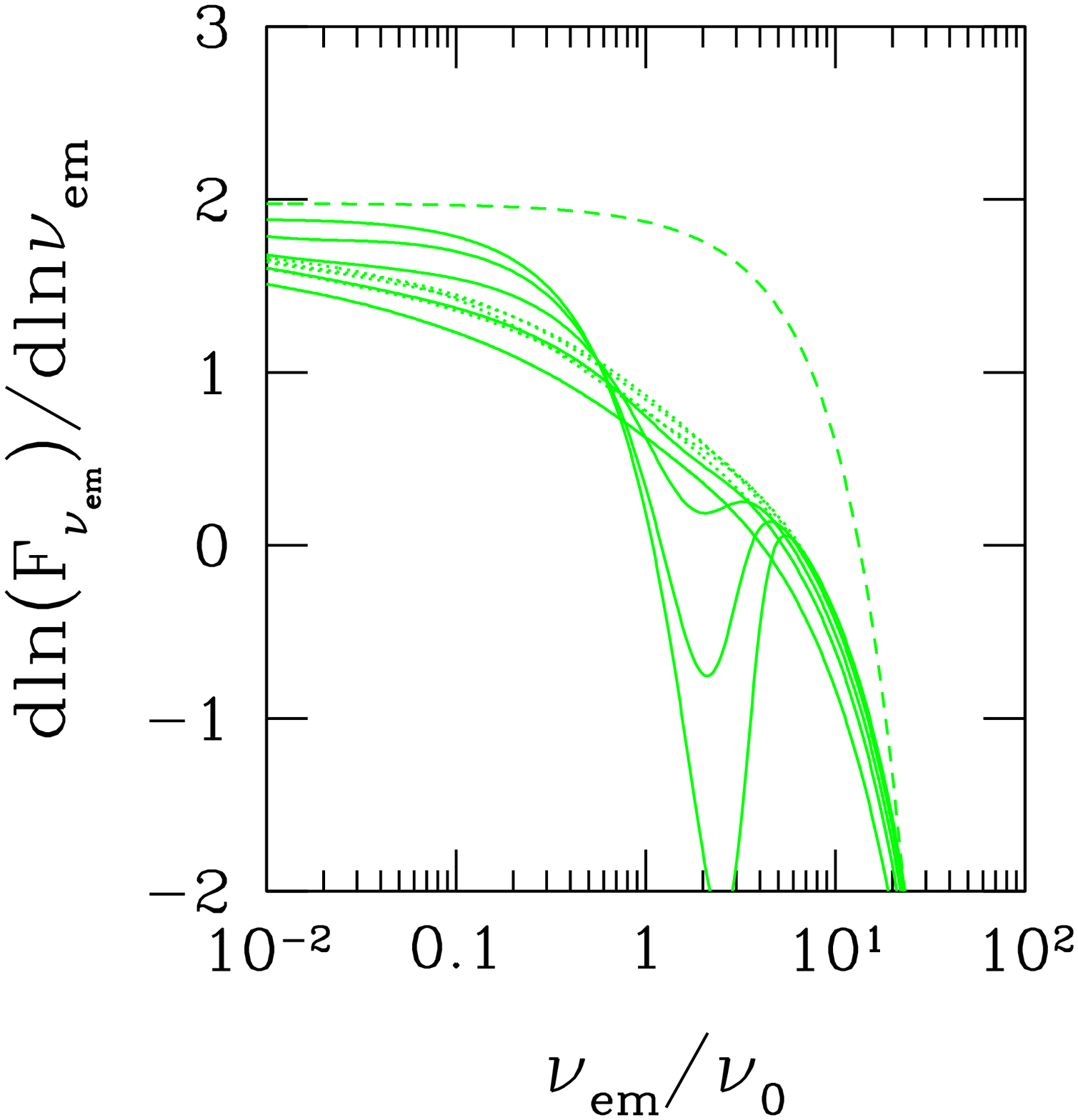}{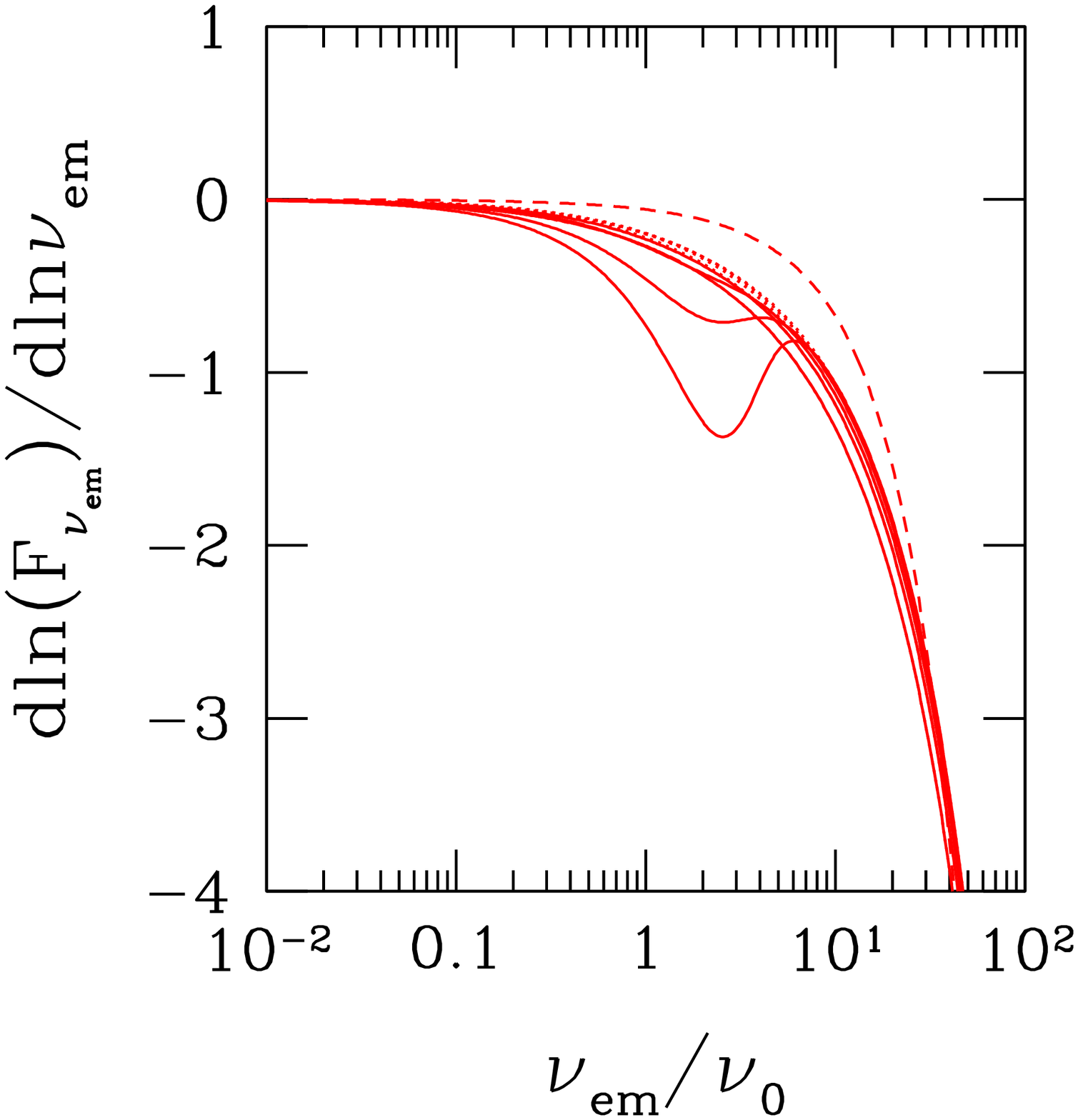}
\caption{\emph{Left panel}:  Slope $d\ln(F_\nu)/d\ln\nu$ of the spectra displayed in Figure \ref{fig:spectra}
(blackbody source), showing a flat spectrum just below the scattered peak, steepening to a Rayleigh-Jeans
slope at low frequencies.  Solid lines:  photon source attenuated by scattering added
to the scattered spectrum.  Dotted line shows blackbody, shifted arbitrarily in peak frequency for comparison.  
\emph{Right panel}: Same outflow, but now a GRB-like photon source.  
Dotted line shows source, shifted in peak frequency.   See text for discussion.}
\label{fig:slope}
\vskip .5in
\end{figure}

\begin{figure}[h]
\centerline{\includegraphics[width=0.55\hsize]{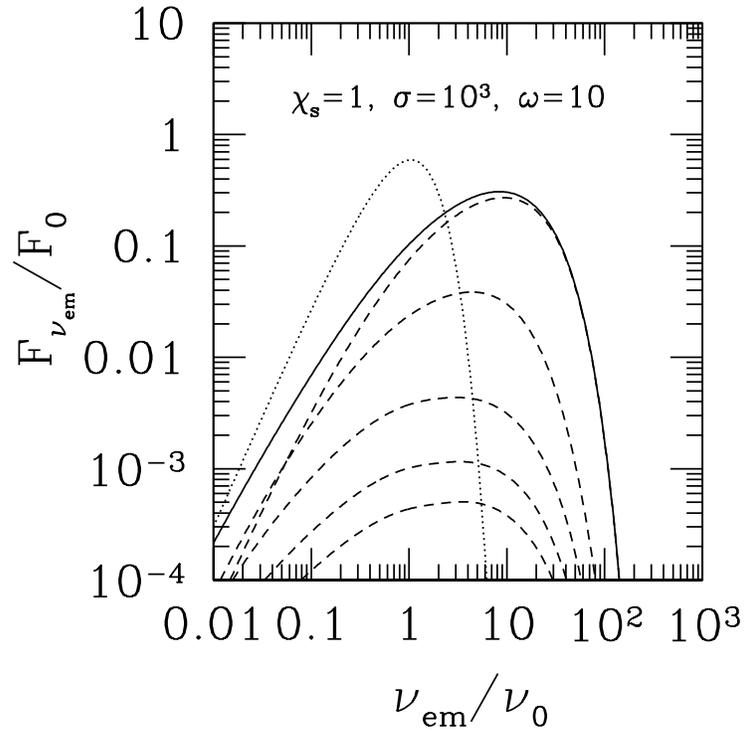}}
\caption{Spectrum of scattered blackbody photons in outflow with $\chi_s = 1$, $\sigma = 10^3$,
$\omega = 10$, and $\tau_{\rm es}(r_s,\mu_{\rm em}=1) = 10$, with the contributions from different radii separated out.
Dashed curves show photons whose radius of last scattering lies in the range $1 < x < 10^{0.5}$, $10^{-0.5} < x < 10$,
$10 < x < 10^{1.5}$, $10^{1.5} < x < 10^2$, $x > 10^2$.   Solid curve: total scattered spectrum.}
\label{fig:spectra3}
\vskip .2in
\end{figure}

In Figures \ref{fig:spectra}-\ref{fig:spectra3} we show spectra calculated using the outflow 
profiles obtained in Section \ref{s:results}, with flow parameters $\sigma=1000$, $\omega=10$.  
A distinct feature of the spectra is a prominent high energy tail appearing in outflows with 
moderate $\chi_s \la 10^3$ -- compare the spectra for $\chi_s = 10^2$ and $10^6$ in Figure \ref{fig:spectra}.
At low compactness, the outflow acceleration is dominated by MHD forces, and $\Gamma \gg \Gamma_{\rm eq}$
at the base of the outflow.   
When the acceleration is dominated by radiation pressure, increasing the optical depth only causes small changes in the
peak of the scattered spectrum (as for $\chi_s = 10^6$).   A direct comparison of outflows with a range of $\chi_s$
is made in Figure \ref{fig:spectra2}.  

The spectral index of the $\chi_s = 10^2$ outflow
is shown in Figure \ref{fig:slope}, for both blackbody and GRB-like seed spectra.
As the optical depth at the base of the outflow increases, the cutoff frequency drops and the spectrum 
softens.  In the blackbody case, the low-energy tail has a fairly constant, Rayleigh-Jeans slope, 
becoming slightly steeper at high compactness, as discussed in Section \ref{s:lowfreq}.

The decomposition of the output spectrum into the components emitted at different radii is shown
in Figure \ref{fig:spectra3}.  The contribution from large radius is in the optically thin regime,
and has a harder spectrum than the (dominant) contribution near the photosphere.
The overall normalization of the spectrum remains essentially constant, $F_{\nuem} \sim F_0$,  
at $\nu_{\rm em} \sim \nu_0$ when $\tau_{\rm es}(r_s,\mu_{\rm min}) \ga 1$.


\subsection{Low-Frequency Spectral Slope}\label{s:lowfreq}

Here we examine in more detail the effect that radiation transfer near a relativistic photosphere will
have on the low-frequency spectrum.  A Rayleigh-Jeans spectral slope arises from side 
scattering a monochromatic source in a locally spherical outflow, and is therefore maintained 
for a black body source (Section \ref{s:bb}).  The optical depth of low-frequency photons scales as
$\tau_{\rm es}(r,\theta_{\rm em}) \propto \theta_{\rm em}^2 \propto \Gamma^{-2}\nu_{\rm em}^{-1}$ 
and emission time $t \propto \nu_{\rm em}^{-1}$ for photons emitted off the axis to the observer.
The increase in optical depth also pushes the off-axis photosphere out 
to a radius $r_\tau(\nu_{\rm em}) \propto \Gamma^{-2}\nu_{\rm em}^{-1}$, where the rate of scatterings in a volume
$\sim r_\tau^3$ is proportional to $(1-\beta)(\Gamma \rho)n_\gamma \propto \Gamma^{-2} r_\tau(\nu_{\rm em})^{-1}
\propto \nu_{\rm em}$.  In the optically thin regime, one therefore finds a low-frequency spectrum
$F_{\nu_{\rm em}} \propto t^{-1} \propto \nu_{\rm em}$, which hardens to $F_{\nu_{\rm em}} \propto
\nu_{\rm em}^2$ if every frequency is emitted from its photosphere.

It is also worth examining briefly how this result for a steady outflow would be modified in the 
case where the outflow is impulsive.  Consider a slightly simpler situation in which the seed photons
flow radially (and are monochromatic).  Then the frequency of the outgoing photon is purely a function of
scattering angle, and we can consider the number of photons scattered the bulk frame of the magnetofluid
in a time interval $dt'$,
\be
d^2N_\gamma = n_{\gamma 0}' c {d\sigma_T\over d\mu_{\rm em}'} d\mu_{\rm em}' dt'.
\ee
Here $n_{\gamma 0}' \sim n_\gamma/\Gamma$ is the bulk frame photon density, and the differential scattering
cross section varies mildly with the rest-frame scattering angle.  Since $\nu_{\rm em}/\nu_0 = 
(1-\beta)/(1-\beta\mu_{\rm em}) = \Gamma^2(1-\beta)(1+\beta\mu_{\rm em}')$, we have
\be
{d^2N_\gamma\over d\tildenuem dt'} = {n_{\gamma 0}' c\over \Gamma^2\beta(1-\beta)} {d\sigma_T\over d\mu_{\rm em}'}.
\ee

The observed arrival time of a photon depends on emission angle and therefore frequency, $dt_{\rm obs} = \Gamma(1-\beta\mu_{\rm em})dt'
= \tildenuem^{-1}\Gamma(1-\beta)dt'$.  But integrating over the entire history of a pulse, the total number of
photons emitted is $dN_\gamma/d\ln\nu_{\rm em} \propto \nu_{\rm em}$.  At very low frequencies, one must compensate
for the expanded photosphere and reduced scattering rate, as outlined above.

\subsection{Semi-analytic Approximation to the Spectrum}

In the Monte Carlo evaluation of the scattered spectrum, we tested a simplified evaluation 
of the optical depth integral, using the small-angle approximation and
assume linear growth of the Lorentz factor, $\Gamma(r_2) = (r_2/r) \Gamma(r) \gg 1$.  
Then equation (\ref{eq:tau0}) becomes
\be\label{eq:tau}
\tau_{\rm es}(r,\mu_{\rm em}) \simeq {\sigma_T\over 6cr}\,{d\dot N\over d\Omega}\,
\left[{1\over\Gamma^2(r)} + \theta_{\rm em}^2\right].
\ee
The output spectra are hard to distinguish from those displayed in Figures (\ref{fig:spectra00}), (\ref{fig:spectra0}).

It is also useful to work out the spectrum of singly-scattered photons in the case where
the seed photons see a small to modest optical depth
Since the matter is cold, it is simplest first to transform into
its rest frame and consider the power scattered into solid angle $d\Omega_{\rm em}'$,
\be
{d^3E'\over d\nu_{\rm em}'d\Omega_{\rm em}'dt'} = 
{3\sigma_T\over 16\pi}\int I_{\nu'}'\left[1+(\hat k'\cdot\hat k_{\rm em}')^2\right]d\Omega'.
\ee
The rest-frame frequency and spectral intensity are $\nu' = \nu_0/\Gamma(1+\beta\mu')$ 
and $I_{\nu'}' = I_\nu/[\Gamma(1+\beta\mu')]^3$.   One sees in Figure \ref{fig:murange} that
the radiation field flows both forward and backward in the bulk frame close to the engine.
At large distances the radiation field continues to collimate even as the $\Gamma$ 
saturates, and $I_{\nu'}'$ is concentrated in the forward (anti-radial) direction.

The power
of the scattered radiation from a single charge is
\ba\label{eq:power}
{d^2E\over d\nu_{\rm em}dt} &=& \int {1\over \Gamma}
{d^3E'\over d\nu_{\rm em}'d\Omega_{\rm em}'dt'} d\Omega_{\rm em}'\nn
&=& {3\sigma_T\over 16\pi}\int d\Omega' d\Omega_{\rm em}' 
{I_0\nu_0\over\Gamma^4(1+\beta\mu')^3}
\delta\left({1+\beta\mu'\over 1+\beta\mu_{\rm em}'}\nu_{\rm em} - \nu_0\right)
\left[1+(\hat k'\cdot\hat k'_{\rm em})^2\right].
\ea
Using
\be
\int d\phi' d\phi'_{\rm em} \left[1+(\hat k'\cdot\hat k_{\rm em}')^2\right]
 = 2\pi^2\left[3 + 3{\mu'}^2{\mu'_{\rm em}}^2 - {\mu'}^2 - {\mu'_{\rm em}}^2\right]
\ee
gives
\be
{d^2E\over d\nu_{\rm em}dt} = {3\pi\sigma_T\over 8}{I_0\nu_0\over\beta\Gamma^4}\tildenuem\int d\mu'
{3 + 3{\mu'}^2{\mu'_{\rm em}}^2 - {\mu'}^2 - {\mu'_{\rm em}}^2\over(1+\beta\mu')^2},
\ee
where $\mu_{em}^{\prime} = \mu_{\rm em}^\prime(\mu',\tildenuem)$ from 
equation (\ref{eq:doppler}),
The range of integration over $\mu'$ is restricted if $\tildenuem > 1$.  Since
$1+\beta\mu' = (1+\beta\mu'_{\rm em})/\tildenuem$, we have
$(1-\beta)/\tildenuem \leq 1+\beta\mu' \leq (1+\beta)/\tildenuem$, and more generally
\be\label{eq:mubounds}
{\rm max}\left[{1-\beta\over\tildenuem},\; {1\over\Gamma^2(1-\beta\mu_{\rm min})}\right] \;\leq\; 
1+\beta\mu' \;\leq\; (1+\beta)\,{\rm min}\left(1,\;{1\over\tildenuem}\right).
\ee

The flux of scattered radiation measured at a large distance is
\be
F_{\nu_{\rm em}}(r) = {1\over r^2}\int^\infty {dr\over \beta(r)c} 
                      {d\dot N\over d\Omega} {d^2E\over d\nu_{\rm em}dt}.
\ee
Normalizing to the incident radiation flux $F_{0}=\pi I_0\nu_0/x^2$, gives
\be\label{eq:fnusing}
{F_{\nu_{\rm em}}\over F_0} =  {3\sigma_T\over 8cr_s}{d\dot N\over d\Omega}\tildenuem
\int d\mu' dx 
{3 + 3{\mu'}^2{\mu'_{\rm em}}^2 - {\mu'}^2 - {\mu'_{\rm em}}^2
\over \Gamma^4\beta^2(1+\beta\mu')^2}.
\ee
Following equation (\ref{eq:tau}), the prefactor can be written as
\be
{3\sigma_T\over 8cr_s}{d\dot N\over d\Omega} \;=\;
    {9\over 4}\Gamma^2(r_s)\;\tau_{\rm es}(r_s,1).
\ee

\section{Summary and Conclusions}\label{s:conclusions}

We have considered a very luminous and strongly magnetized outflow outside its scattering photosphere.  
The outflow is accelerated to a high Lorentz factor by a combination of the Lorentz force (which acts in a cold MHD flow)
and the radiation scattering force (dominant in thermal fireballs).   A range of radiation intensities is considered,
extending from an almost cold flow to one in which the radiation and magnetic Poynting fluxes are comparable.
The calculations described in this first paper assume that the poloidal magnetic field is strictly monopolar,
which results in a near degeneracy between magnetic pressure gradient and curvature forces.  Similar solutions will
obtain for any part of an MHD outflow in which the magnetic flux surfaces are unfavorably curved and the Lorentz
force remains weak.   The opposing case, corresponding to a flared jet that breaks out of a confining medium, is 
examined in detail in paper II.  

The radiation force dominates the acceleration if the compactness $\chi \ga \sigma$ at the photosphere.  This inequality is
easily satisfied if the outflow is optically thick near the engine:  one has $\chi(r_\tau) \sim 6\Gamma^2(r_\tau)\sigma$
in an outflow with comparable Poynting and photon energy fluxes.  Radiative driving is especially efficient beyond
the fast critical point, even though the magnetic field dominates the inertia of the outflow.  The solutions we obtain for 
high radiation intensities can easily be rescaled to an outflow with relativistic bulk motion at a displaced photosphere:
then the flow profile $\Gamma(r)$ is linear in the inner parts of the outflow at both large and small scattering depths.

We have considered the imprint of bulk Compton scattering on a photon seed with an exponential, high-frequency spectral cutoff. 
The spectrum is strongly modified when the radiation compactness is low enough that outward acceleration
is dominated by the Lorentz force {\it and} the seed photon beam is wider than the Lorentz cone of the
magnetofluid.  In this situation, the magnetofluid is pushed quickly to a high Lorentz factor outside
its photosphere, where it feels a strong photon drag.   Then the output spectrum
extends above the seed peak frequency, with its low-frequency part depending on the shape of the seed.
In the case of a blackbody seed, the spectral slope in between the seed thermal peak and the scattered peak
is softer than Rayleigh-Jeans, but harder than is typical of GRBs at low frequencies.

We have also considered a seed spectrum $F_\nu \sim {\rm const}\times e^{-h\nu/kT_0}$, representing the low-frequency 
part of the Band function absent the high-frequency tail.  In that case, the scattered spectrum extends the flat portion of the seed spectrum upward
in frequency.  For both types of seed spectrum, the residual amplitude of the seed thermal
peak that persists in the transmitted spectrum depends on the optical depth.  A seed photon beam that is much wider
than the Lorentz cone of the magnetofluid sees a large optical depth (compared with the optical depth of the 
more strongly beamed scattered photons), which means that the transmitted thermal peak is relatively
weak.  In the context of GRBs, the angular broadening of the seed photons could result from scattering by
a second, slower component of the outflow that is swept up at the head of the jet \citep{thompson06}.

In principle, no fine tuning of the optical depth surface is needed to make radiative and MHD acceleration
competitive near the photosphere.  The effective magnetization is much reduced below the photosphere,
where the stress-energy of the photons couples to the matter and thence to the magnetic field:
$\sigma_{\rm eff} \sim L_{\rm P}/L_\gamma$.  Once the radiation begins to stream freely, the 
outflow experiences both a rapid increase in magnetization, and a strong outward force from the self-collimating
radiation field.  This effect is examined in detail in Paper II.

\subsection{Connection with GRBs}

At first sight, one might associate the high-energy tail of the scattered photon spectrum with the observed 
high-energy tails of GRBs, and the seed photon energy with the observed spectral peak energy $E_{\rm pk}$.
But the calculated spectrum is relatively hard compared with the high-energy tails of GRBs, and it is limited in spectral width.

Instead it appears more promising to identify the high-energy peak of the scattered spectrum with the measured $E_{\rm pk}$ --
at least in some bursts or possibly some phases of the burst emission.
Then an additional source of dissipation, which is left out of our calculations, is needed to generate the high-energy tail.
The seed thermal photons, generated deep in the outflow, provide a buffer that suppresses bursts with low $E_{\rm pk}$, 
but upscattering allows a range of {\it higher} $E_{\rm pk}$ values.   
We leave open here the nature and origin of the low-frequency seed spectrum, except to say that a hard Rayleigh-Jeans
slope is by no means guaranteed if thermalization occurs at an intermediate optical depth. 

In this context, it is worth recalling some features of 
the \cite{amati02} relation between $E_{\rm pk}$ and the apparent isotropic energy $E_{\rm iso}$ of GRBs.  This relation 
is suggestive of jet breakout from the core of a Wolf-Rayet star (e.g. \citealt{thompson06}), but it appears to represent 
a boundary in the $E_{\rm pk}$-$E_{\rm iso}$ plane.   The observed bursts have $E_{\rm pk}$ lying 
on or above the Amati et al. line, with a strong deficit mainly below the line:  for example, a significant proportion
of BATSE burst spectra are too hard to be consistent with this relation, independent of the (unknown) redshift
\citep{nakar05}.

Evidence for a low frequency blackbody component has been found in some gamma-ray bursts (e.g. \citealt{axelsson12}).  It is natural
to try to accomodate such components in the present calculation, where the thermal seed should be partially transmitted.
The measured soft blackbody component sits a factor $\sim 1/30$ below the burst peak, which requires a large broadening
of the seed photons (by a factor $\sim \Gamma \theta_{\rm seed} \sim 5$ in angle, where $\theta_{\rm seed}$ is the angular width
of the seed component).  If the decrease in the peak frequency during the burst evolution were due {\it mainly} to a drop
in $\Gamma \theta_{\rm seed}$, then one would expect to see a strong re-emergence of the seed thermal peak alongside
the scattered peak, which is generally not observed.  When considering the emission from relativistic outflows, one
must always keep in mind the basic degeneracy between temporal and spectral degrees of freedom, and the angular degree
of freedom.  Therefore the presence of a soft thermal component in the spectrum may signal the presence of off-axis
emission that does not, necessarily, interact with the emission zone of the harder component.  

Finally, onee should keep in mind
that the scattered thermal radiation emitted by an electromagnetic outflow could represent a subcomponent of the burst emission, 
with some other non-thermal (e.g. synchrotron) process dominating \citep{peer06,zhang11}.  In that case, the photon field would 
have a smaller (but not necessarily negligible) influence on the outflow acceleration.

\subsection{Alternative Acceleration Mechanisms for Strictly Radial Magnetized Outflows}

We find that a hot electromagnetic outflow typically experiences a strong radiation force before it expands far enough
that radial inhomogeneities become important.
The acceleration of a static, bounded magnetic slab, studied by \cite{granot11}, could be relevant for the later
stages of impulsive GRB outflows, outside a radius $\sim c\Delta t$.  But if a magnetized shell already moves relativistically 
at this radius, only a thin outer layer, comprising a fraction $\sim 1/2\Gamma^2$ of the shell, would experience a strong outward magnetic pressure gradient
force.  Its interaction with slower material, swept up from a Wolf-Rayet star or a preceding neutron-rich
wind, must then be taken into account.  

Other acceleration mechanisms have been suggested which depend on more complicated, non-ideal MHD effects, such
as the creation of a (net) outward pressure gradient force by reconnection of a toroidal magnetic field 
\citep{drenkhahn02}.  Zones of alternating $B_\phi$, separated by current sheets, are indeed present 
in the force-free solution to the oblique rotator \citep{spitkovsky06}, in a
zone straddling
the rotational equator.   An active dynamo operating in a GRB engine could also lead to stochastic reversals
in the wind magnetic field \citep{thompson06}.  But when the increase in flow inertia associated with
particle heating and radiation is taken into account, we have argued that cancelling even half the magnetic flux
leads only to mildly relativistic radial motion.  Magnetic reconnection plays
a more natural role in the GRB phenomenon by modifying the gamma-ray spectrum, via particle heating and stochastic bulk motions.

\acknowledgements
We thank the NSERC of Canada for financial support, and the referee for comments.

\appendix
\section{Rotating Emission Surface}\label{s:rotation}

The photon source rotates rapidly in some cases, e.g. a rapidly rotating
star such as a millisecond magnetar, or the merged remnant of a white dwarf binary.  
We can approximate the effect of a rotating emission surface by setting
\be\label{eq:betarad}
\beta_{\phi}\rightarrow
\beta_{\phi}-\frac{\beta_{\phi,R}}{x}
\ee  
in equations (\ref{e:R}), and (\ref{e:P}).  Here $\beta_{\phi,R}$ is
a constant representing the aberration of the outflowing photons at
$r = r_s$ ($x = 1$).
In this situation, plasma near the emission surface can more 
easily co-rotate with the radiation field while still being accelerated outward.

\begin{figure}[h]
\centerline{\includegraphics[width=0.6\hsize]{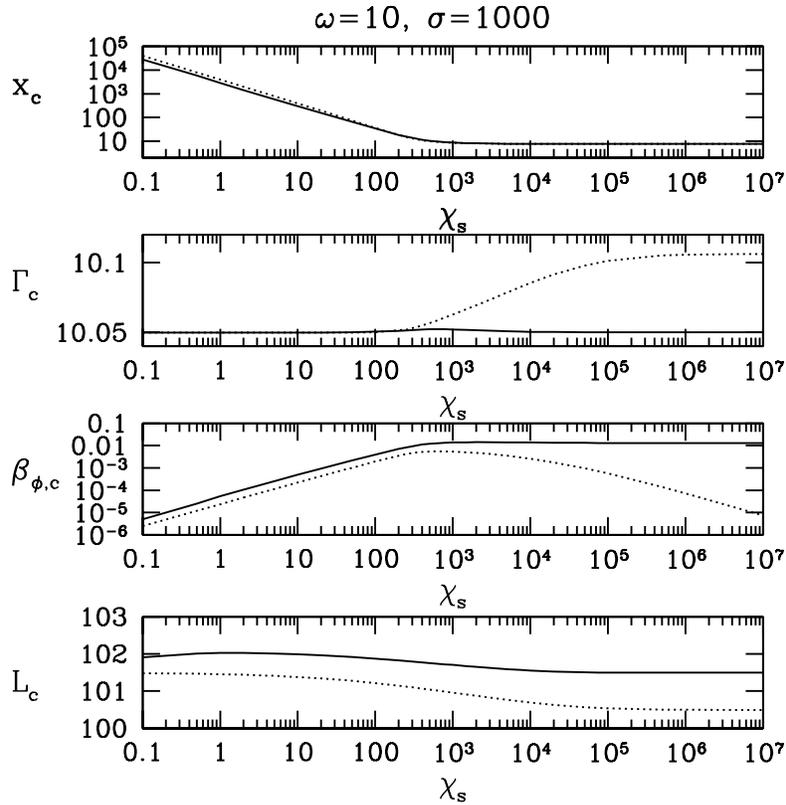}}
\caption{Effect of rotation of the photon source (compactness $\chi_s$) on the properties of the fast critical point:
radius $x_c$ and flow variables $\Gamma$, $\beta_{\phi}$ and $\mathcal{L}$.   Solid line: $\beta_{\phi,R}=1/\omega
= 0.1$ [equation (\ref{eq:betarad})]; dotted line: $\beta_{\phi,R}=0$.  Flow magnetization $\sigma = 10^3$.}
\label{fig:criticalpointsAF}
\vskip .2in
\end{figure}

The value of $\beta_{\phi,R}$ depends on the type of source.  One
has $\beta_{\phi,R} \sim \Omega r_s/c \equiv \omega $ when the photons flow from
the surface of a star of radius $r_s$ through a transparent wind.
On the other hand, if the outflow is optically thick in a narrow radial zone
close to the engine, then one expects $\beta_{\phi,R} \sim (\Omega r_s/c)^{-1}
\sim \omega^{-1}$ based on the conservation of angular momentum from the
light cylinder out to the transparency surface ($x = 1$).  

In a first approximation, rotation of the photon source makes only small
changes to the profiles of Lorentz factor and angular momentum in the outflow.
For completeness, we discuss some of the detailed changes that do result in
the flow parameters at the fast point.  These effects are largest in intense
radiation fields with compactness $\chi_s>4\sigma^{4/3}/\omega^2$. 
Then, as in our previous calculations with a non-rotating photon source (Section \ref{s:largechi}), 
the location of the critical point is determined by setting $\Gamma_c = \Gamma_{\rm eq}(r)$
[equation (\ref{eq:gameq})].  Corotation of the fluid and the radiation field implies
\be
\beta_{\phi,c} \simeq \frac{\beta_{\phi,R}}{x_c},
\ee 
where $x_c\simeq \Gamma_c/3^{1/4}$.
When $\beta_{\phi}\neq0$, the fast speed (\ref{e:msspeed}) becomes 
\be
u^{3}\simeq
\sigma\left[1+\Gamma^{2}\frac{B_{r}^{2}}{B_{\phi}^{2}}\left(1-\beta_{\phi}x\omega\right)^2\right]\simeq   
\sigma.
\ee
At large compactness, the Lorentz factor and angular momentum at the critical 
point are given by 
\be
\Gamma_c \simeq
\sigma^{1/3}\left(1+\sqrt{3}(\omega^{-1}-\beta_{\phi,R})^{2}\right)^{1/3};
\quad \quad
\mathcal{L}_c\simeq
\frac{\Gamma_c}{\omega}\left[\beta_{\phi,R}\omega+\frac{\sigma^{2/3}}{\left(1+\sqrt{3}(\omega^{-1}-\beta_{\phi,R})^2\right)^{1/3}}\right]
\quad\quad({\rm large~\chi_s}).
\ee
One sees that the inertia added by the radial magnetic field can be significantly
reduced when $\beta_\phi \simeq 1/x\omega$, that is when $\beta_{\phi,R} \simeq \omega^{-1}$.

A comparison of flow profiles with, and without, rotation of the photon
source is made in Figure \ref{fig:criticalpointsAF}.  There is little
change in the critical point radius, but $\Gamma_c$ remains close
to $\sigma^{1/3}$ at all values of the compactness, since we have taken $\beta_{\phi,R} \simeq \omega^{-1}$.
The magnetofluid rotates more rapidly near the critical point at large values of $\chi_s$.



\end{document}